\documentclass[showpacs,preprintnumbers,amsmath,amssymb,11pt]{article}
\usepackage[utf8]{inputenc}
\usepackage[english]{babel}
\usepackage{amsmath}
\usepackage{graphicx}
\usepackage[colorinlistoftodos]{todonotes}
\usepackage{caption}
\usepackage{subcaption}
\usepackage{tabularx}
\usepackage{bbm}
\usepackage[normalem]{ulem}
\usepackage{listings}
\usepackage{amsthm}
\usepackage{algorithm}
\usepackage{algpseudocode}
\usepackage{mathtools}
\usepackage{amsmath}
\usepackage[toc,page]{appendix}
\usepackage{booktabs}
\usepackage{xcolor}
\usepackage{amsfonts}
\usepackage{epstopdf}
\usepackage{adjustbox}
\usepackage{csquotes}
\usepackage{enumitem} 
\usepackage{makecell}
\usepackage{tikz}

\usetikzlibrary{decorations.pathreplacing}

\usetikzlibrary{arrows}
\usetikzlibrary{shapes}

\usepackage[numbers]{natbib}

\DeclareMathOperator*{\argmin}{arg\,min}

\makeatletter
\def\BState{\State\hskip-\ALG@thistlm}
\makeatother

\definecolor{darkgreen}{rgb}{0,0.5,0}
\definecolor{darkred}{rgb}{0.5,0,0}

\usepackage{amsmath,amsfonts,amssymb,amsthm,epsfig,epstopdf,titling,url,array}

\theoremstyle{plain}
\newtheorem{thm}{Theorem}[section]

\newtheorem{prop}[thm]{Proposition}

\theoremstyle{plain}
\newtheorem{defn}{Definition}[section]

\theoremstyle{remark}
\newtheorem{rem}{Remark}

\newcommand{\E}{\mathbb{E}}
\newcommand{\pdf}{\textit{pdf}}
\newcommand{\chf}{\textit{chf}}

\newcommand{\R}{\mathbb{R}}
\newcommand{\N}{\mathbb{N}}

\newcommand{\rv}{\textit{rv}}
\newcommand{\id}{\textit{id}}
\newcommand{\iid}{\textit{iid}}
\newcommand{\sd}{\textit{sd}}

\newcommand{\eqd}{\stackrel{d}{=}}

\newcommand{\arem}{$a$-remainder}
\newcommand{\VGpp}{VG++}

\newcommand{\norm}{\mathcal{N}}

\newcommand{\poiss}{\mathcal{P}}
\newcommand{\pol}{\overline{\mathfrak{B}}}

\newcommand{\unif}{\mathcal{U}}

\newcommand{\erl}{\mathcal{E}}

\newcommand{\Levy}{L\'{e}vy}

\newcommand\Beta{\mathrm{B}}

\newcommand{\MLE}{\emph{MLE}}
\newcommand{\GMM}{\emph{GMM}}
\newcommand{\NLLS}{\emph{NLLS}}
\newcommand{\zapp}{Z_a^{++}}

\title{The Variance Gamma++ Process and \\
Applications to Energy Markets
}

\author{
	Matteo Gardini\thanks{Department of Mathematics, University of Genoa, Via Dodecaneso 16146, Genoa, Italy, email gardini@dima.unige.it}
	\and
	Piergiacomo Sabino\thanks{Quantitative Risk Management, E.ON SE,
		Br\"usseler Platz 1, 45131 Essen, Germany, email piergiacomo.sabino@eon.com}
	\and
	 Emanuela Sasso\thanks{Department of Mathematics, University of Genoa, Via Dodecaneso 16146, Genoa, Italy, email sasso@dima.unige.it}}

\date{\today}
\begin{document}
\maketitle

\begin{abstract}
\noindent The purpose of this article is to introduce a new \Levy\ process, termed  Variance Gamma++ process, to model the dynamic of assets in illiquid markets. Such a process has the mathematical tractability of the Variance Gamma process and is obtained applying the  self-decomposability of the gamma law. Compared to the Variance Gamma model, it has an additional parameter representing the measure of the  trading activity. We give a full characterization of the Variance Gamma++ process in terms of its characteristic triplet, characteristic function and transition density. In addition, we provide efficient path simulation algorithms, both forward and backward in time. We also obtain an efficient \enquote{integral-free} explicit pricing formula for European options. These  results are instrumental to apply Fourier-based option pricing and maximum likelihood techniques for the parameter estimation.
Finally, we apply our model to illiquid markets, namely to the calibration of European power future market data. We accordingly evaluate exotic derivatives using the Monte Carlo method and  compare these values to those obtained  using the Variance Gamma process and give an economic interpretation of the obtained results. Finally, we illustrate an extension to the multivariate framework.
\end{abstract}

\section{Introduction}

The purpose of this study is to introduce  a new \Levy\ process related to the Variance Gamma process which inherits its mathematical tractability and financial interpretation. It has only an additional parameter which measures the trading activity and therefore the liquidity regime. We call such a new process Variance Gamma++ (\VGpp). 

\par Models based on the Variance Gamma distribution are widely used in finance since the introduction of the Variance Gamma process by \citet{MadanSeneta90}. Such a process presents many interesting properties: both characteristic function and density are available in a closed form and, moreover, a closed formula for European options  is known. Finally, efficient methods for path simulations can be used in order to simulate the process and hence to price exotic contingent claims. All these properties together with the fact that the model overcomes some well known limits of the \citet{BLS1973} model, make it a good candidate for financial markets modeling. 

In contrast to the classical \citet{BLS1973} market, where real data description is based on the standard Brownian diffusion-type processes, the Variance Gamma assumes that dynamics of the price or of the returns depends on a time-changed Brownian motion where the time-change is given by a gamma process. Such a random time process, called subordinator, can be interpreted as trading activity, in the sense that the price does not evolve in terms of the physical time but instead in terms of the random transactions exchanged in the market. 

This interpretation has been explored using different types of subordinator processes, for instance \citet{BN98} takes an Inverse Gaussian  subordinator and also the CGMY model, introduced in \citet{CGMY2002} which generalizes the Variance Gamma model, under some parameter constrains can be seen a time-changed Brownian motion. All these models are pure jump models with infinite activity that differ from jump-diffusion models (see for instance \citet{Merton76} and \citet{Kou2002}) where the jumps are interpreted as sudden news in the market.  

However, some real data exhibit characteristic periods of constant values especially in illiquid markets like some not so mature energy markets. In such cases, adopting the financial interpretation that the subordinating process represents the trading activity, the gamma process (and the other subordinators mentioned above) imply that in any finite time-interval the number of trades cannot be zero because its trajectory is strictly increasing.
The Variance Gamma process essentially exhibits an infinite number of jumps in any finite time interval and hence its trajectories can not be constant over time (see \citet[Lemma ~2.1]{CT2003}). Market liquidity is generally strictly related to the amount of registered transactions between counterparts. Therefore, a zero variation of the price over the time period $\Delta t$ usually appears when no market transactions occur.  
 
The main idea of this research is to replace the gamma process by another process related to it, which may be constant in time and keeps the right properties to still behave as a subordinator. The new subordinator is then of finite activity and the probability of having no transactions in a finite period of time will not be null. 

\par To this end we use the well-know self-decomposabilty of the gamma law (see \citet{Gri03}). We recall that a random variable $X$ is said  to have a self-decomposable law (see \citet{Sato} and \citet{Cufaro08}), if for all $a \in \left(0,1\right)$ there exist two independent random variables $Y$ and $Z_{a}$ such that $X\eqd Y$ and:
\begin{equation*}
	X \eqd aY + Z_{a}.
\end{equation*}
In the following we will refer to $Z_{a}$ as the \arem\ of the \sd\ law. It turns out that the law of $Z_{a}$ is infinitely divisible (see \citet{Sato}) and  one can construct the associated \Levy\ process $Z_{a}^{++}=\left\{Z_{a}^{++}(t);t\ge 0\right\}$. 

Our approach consists in taking the subordinator $Z_{a}^{++}$, from the \arem\ of the gamma law to construct the new \VGpp\ process $X=\left\{X(t); t\ \ge 0\right\} = \left\{W(Z_{a}(t));t \ge 0\right\}$  where $W=\left\{W(t);t \ge 0\right\}$ is a Brownian motion with drift $\theta\in \R$ and diffusion $\sigma >0$ independent from $Z_{a}^{++}$.
Denoting with $X$ the log-price process of a risky asset and $\Delta X = X(t+\Delta t) - X(t)$ its increment over the time interval $\Delta t$, we show that $\mathbb{P}\left(\Delta X = 0\right)>0$, therefore we have a non zero probability to have no transactions in the time interval $\Delta t$. 
In particular, we show that the parameter $a$ plays the role of an indicator of the trading activity. 

\par Accordingly, we derive the \Levy\ measure, the transition density and the characteristic function  in closed form. However, the new process has finite activity, but can also be written as the difference of two independent subordinators and keeps the mathematical tractability of the Variance Gamma process. As a consequence, we obtain a closed formula for European call options,  which is an infinite weighted sum of call options priced under the Variance Gamma model, where the shape parameter of the underlying gamma subordinator is now an integer. Such a formula does not require any numerical integration, but can be reduced to matrix multiplications which are faster than numerical integration algorithms.

\par The paper is structured as follows: in Section \ref{sec:notationandpreliminary} we introduce some notation and we give some preliminary remarks which are instrumental to give the full characterization of the \VGpp\ process: moreover, we introduce two different algorithms to simulate the skeleton of the process $Z_{a}^{++}$. In Section \ref{sec:VG++process} we study the mathematical properties of the \VGpp\ process: we give its full characterization in terms of its \Levy\ triplet, we derive a close formula for European call options pricing and, finally, we derive the law of its \Levy\ bridge. This last result is then used to develop an efficient method to simulate the \VGpp\ process backward in time. Section \ref{sec:financialapplications} illustrates numerical results and a possible financial application. In Subsection \ref{subsec:optionpricingmethods} we compare the option pricing using the FFT method proposed by \citet{Carr99} and by the Monte Carlo scheme   with that obtained using the closed formula presented in Section \ref{sec:VG++process}. In Subsection \ref{subsec:Calibration} we calibrate the \VGpp\ model on power energy future markets, in Subsection \ref{subsec:pricing} we compare the prices of exotic derivatives obtained using the Variance Gamma model and the \VGpp\ model. Finally, in Section \ref{sec:multivariateframework} we briefly discuss how to extend the model to a multivariate framework and Section \ref{sec:conclusions} concludes and gives some insights about possible future inquires.

\section{Notation and preliminary remarks}
\label{sec:notationandpreliminary}
In this section we introduce the notation and the shortcuts that will be used
throughout the paper and present some concepts and instrumental results for the  construction of the \VGpp\ process.

\subsection{Notation}\label{sec:notation}
We write $\Gamma(\alpha, \beta)$ to denote the gamma law with scale $\alpha>0$ and rate $\beta>0$. Of course, when $\alpha=n\in\N$, such a law coincides with the Erlang distribution denoted $\erl_{n}(\beta)$, for simplicity we drop $n=1$ for the exponential distribution. We write $\poiss(\lambda)$ to denote the Poisson law with parameter $\lambda>0$, $\norm(\mu, \sigma)$ to denote the Gaussian distribution with mean $\mu\in\R$ and variance $\sigma^2>0$. Moreover, we write $\unif([0, 1])$ to denote the uniform distribution in $[0, 1]$. We use the shortcuts \id\ and \sd\ for infinitely divisible and self-decomposable distributions, respectively. We use the shortcut \rv\ for random variable and \iid\ for independently and identically distributed, whereas we use
\chf\ and \pdf\ as shortcuts for characteristic function and density function, respectively.

\subsection{Preliminary remarks}\label{sec:preliminaries}

A \rv\ $X$ is said to have a \sd\ law if for all $a\in \left(0,1\right)$ there exist a \rv\ $Y$ with the same law of $X$ and a \rv\ $Z_{a}$ independent of $Y$ such that
\begin{equation*}
	X \stackrel{d}{=} aY + Z_{a}.
\end{equation*}	
In the following we will refer to $Z_{a}$ as the \arem\ of the \sd\ law.
If we denote by $\phi_{X}\left(u\right)$ the \chf\ of $X$ and by $\phi_{Z_{a}}\left(u\right)$ the \chf\ of $Z_{a}$ we have that:
\begin{equation}
	\phi_{X}\left(u\right)= \phi_{X}\left(au\right)\phi_{Z_{a}}\left(u\right).
	\label{eqn:chfsdrelation}
\end{equation} 
It can be shown that the law of the \arem\ of a \sd\ law is \id\ (see \citet[Proposition~15.5]{Sato}).
On the other hand, it is well-known that the gamma law is \sd\ (see \citet{Gri03}) and hence the law of its \arem\ $Z_{a}$ is also \id.  

\begin{defn}
 We say that $Z_{a}$ has a gamma++ law, and we write $Z_{a} \sim \Gamma^{++}\left(a, \alpha,\beta\right)$, if $Z_a$ is the \arem\ of a $\Gamma(\alpha, \beta)$  distribution. 
\end{defn}
By Equation \eqref{eqn:chfsdrelation} it follows that the \chf\ of $Z_a$ is	
\begin{equation}
		\phi_{Z_{a}}\left(u\right) = \left(\frac{\beta-iua}{\beta -iu}\right)^{\alpha}.
		\label{eqn:chfgammareminder}
\end{equation}
In particular its mean the variance are given by:
\begin{equation*}
	\E\left[Z_{a}\right] = \left(1-a\right)\frac{\alpha}{\beta}, \qquad Var\left[Z_{a}\right] = \left(1-a^{2}\right)\frac{\alpha}{\beta^2}.
\end{equation*}

Based on the observations above and the findings of \citet{SabinoPetroni2020GammaRelated}, in this section we  construct the \Levy\ process $\zapp = \left\{\zapp(t);t \ge 0\right\}$ associated to the law of the \arem\ of the gamma law, e.g. $\zapp(1)\eqd Z_a$. To this end, we recall the following known results (see \citet{SabinoPetroni2020GammaRelated} for details and proofs).

\begin{defn}
	A discrete \rv\ $S$ is said to be Polya distributed, $S\sim \pol\left(\alpha,p\right)$, with parameters $\alpha >0$ and $p \in \left(0,1\right)$, if its probability mass function has the following form:
	\begin{equation*}
		\mathbb{P}\left(\left\{S=k\right\}\right) =\binom{\alpha +k-1}{k} \left(1-p\right)^{\alpha}p^{k},\qquad k =0,1,\dots
	\end{equation*}
	where:
	\begin{equation*}
		\binom{\alpha}{k} = \frac{\alpha \left(\alpha-1\right)\dots \left(\alpha-k+1\right)}{k!}, \qquad \binom{\alpha}{0} = 1.
	\end{equation*}
\end{defn}

\begin{prop}
	\label{prop:ZaRapresentation}
	 Consider $X\sim \Gamma\left(\alpha,\beta\right)$, then 
	$$
	Z_{a}\stackrel{d}{=} \left\{\begin{array}{ll}\sum_{i=1}^{S} X_i, &\mbox{when $S>0$}\\
		0, &\mbox{when 
			$S=0$}\end{array}\right.$$
	when $X_i\sim \erl(\beta/a)$ is a sequence of \iid\ \rv's and $S \sim \pol\left({\alpha,1-a}\right)$. In particular ${Z_a}_{|S=s}\sim \Gamma (s, \beta/a)$, when $s>0$.
\end{prop}
%
%
%

\begin{prop}
	The \pdf\ $g_{a}\left(x\right)$ of $Z_{a} \sim \Gamma^{++}\left(\alpha,\beta\right)$ is given by:
	\begin{equation}
		g_{a}\left(x\right)= a^{\alpha}\delta_{0}\left(x\right) + \sum_{n \ge 1} \binom{\alpha + n -1}{n} a^{\alpha} \left(1-a\right)^{n}f_{n,\beta/a}\left(x\right) \mathbbm{1}_{\left(0,\infty\right)}\left(x\right)dx
		\label{eqn:Zadensity}
	\end{equation}
	where $\delta_{0}\left(x\right)$ is the Dirac function, $f_{n,\beta/a}\left(x\right)$ is the \pdf\ of an Erlang law with parameters $n$ and $\beta /a$ which is given by:
	\begin{equation*}
		f_{n,\beta/a}\left(x\right) = \left(\frac{\beta}{a}\right)^{n}\frac{x^{n-1}e^{-\beta\,x/a}}{\left(n-1\right)!}\mathbbm{1}_{\left[0,\infty\right)}\left(x\right).
	\end{equation*}	
	\label{prop:pdfofZ}
\end{prop}

We remark that the law of $Z_{a}$ can be seen as a mixture of Erlang laws with parameter $\beta/a$, where the mixing distribution is a Polya distribution, plus a degenerate law at $x=0$.

From Corollary \ref{prop:ZaRapresentation} we can define the process $\zapp$ as follows:
\begin{equation}
	\zapp(t)\eqd \left\{\begin{array}{ll}\sum_{i=1}^{S(t)} X_i, &\mbox{when $S(t)>0$},\\
		0, &\mbox{when 
			$S(t)=0$}\end{array},\right.
		\label{eqn:ZaProcess}
\end{equation}
where $X_i\sim \erl(\beta/a)$ is a sequence of \iid\ \rv's and $S = \left\{S(t);t \ge 0\right\}$ is a Polya process such that for each $t\ge 0$, $S(t) \sim \pol\left({\alpha t,1-a}\right)$. The construction is mathematically consistent since the Polya distribution is \sd\ and therefore the Polya process is a \Levy\ process. \\

\par We proceed then in the derivation of the characteristic \Levy\ triplet of the process $\zapp$. 
We rely on the the following proposition proven in \citet{petronisabino2020} that relates the characteristic triplet of a \sd\ law with that of its \arem.

\begin{prop}
	\label{prop:characteristictripletarem}
	Consider a \sd\ law with \Levy\ triplet $\left(\gamma,\sigma, \nu\right)$, where $\sigma>0$ is the diffusion and $\nu$ is the \Levy\ measure. Then for every $a\in \left(0,1\right)$ the law of its \arem\ has \Levy\ triplet $\left(\gamma_a,\sigma_a, \nu_a\right)$:
	\begin{align*}
		\gamma_{a} & = \gamma\left(1-a\right) - a\int_{\R} sign\left(x\right) \left(\mathbbm{1}_{|x|\le 1/a} -\mathbbm{1}_{|x|\le 1}\right)|x|\,\nu(x)\,dx, \\
		\sigma_{a} & = \sigma \sqrt{1-a^2}, \\
		\nu_{a}\left(x\right) & = \nu\left(x\right) - \frac{\nu\left(x/a\right)}{a}.
	\end{align*}
\end{prop}

\begin{prop}
	Consider the process $\zapp$, then 
	\begin{enumerate}[label=(\roman*)]
			\item The characteristic triplet $\left(\gamma_{a},\sigma_{a},\nu_{a}\right)$ of $\zapp$ is given by:
		\begin{align*}
			&\gamma_{a}  = \left(1-e^{-\beta}\right)-a\left(1-e^{-\beta/a}\right),\\
			&\sigma_{a} = 0,\\
			&\nu_{a}\left(x\right) = \frac{\alpha}{x}\left(e^{-\beta x} - e^{-\beta x/a}\right)\mathbbm{1}_{\left(0,\infty\right)}\left(x\right).
		\end{align*}
		\item $\zapp$ has finite variation and, in particular, is a subordinator.
		\item $\zapp$ has finite activity and therefore it is a compound Poisson process with intensity $\lambda= \alpha \log\left(1/a\right)$ and the distribution of the jumps $f\left(x\right)$ is given by:
		\begin{equation*}
			f\left(x\right) = \int_{1}^{1/a} \frac{1}{y\log\left(1/a\right)}\cdot \beta y e^{-\beta x y}dy.
		\end{equation*}
	\end{enumerate}
\label{prop:chtripletteZa}
\end{prop}

\begin{proof}
	
	\begin{enumerate}[label=(\roman*)]
		\item As a direct consequence of Proposition \ref{prop:characteristictripletarem}. 
		\begin{align*}
			\sigma_{a} &= 0,	\\
			\nu_{a}\left(x\right) &= \frac{\alpha}{x}e^{-\beta x}\mathbbm{1}_{\left(x,\infty\right)}\left(x\right) - \frac{1}{a}\left(a \cdot\frac{\alpha}{x}e^{-\beta x/a}\right)\mathbbm{1}_{\left(x,\infty\right)}\left(x\right) = \frac{\alpha}{x}\left(e^{-\beta x} - e^{-\beta x/a}\right)\mathbbm{1}_{\left(x,\infty\right)}\left(x\right)
		\end{align*}
		\begin{equation*}
			\begin{split}
				\gamma_{a} &= \gamma\left(1-a\right) - a\int_{0}^{1/a}\alpha  e^{-\beta x}dt + a\int_{0}^{1} \alpha e^{-\beta x}dx \\
				&= \gamma\left(1-a\right) -a\alpha  \left(\left.-\frac{e^{-\beta x}}{\beta}\right]_{0}^{1/a} + \left. \frac{e^{-\beta x}}{\beta} \right]_{0}^{1}\right) \\
				&= \gamma\left(1-a\right) +  \frac{a\alpha }{\beta}\left(e^{\beta /a} - e^{-\beta}\right) \\
				& = \frac{\alpha }{\beta}\left(1-e^{-\beta}\right)\left(1-a\right) +  \frac{a\alpha }{\beta}\left(e^{\beta /a} - e^{-\beta}\right) \\
				& = \frac{\alpha }{\beta} \left(1-e^{-\beta} - a\left(1-e^{-\beta/a}\right)\right) \\
				& = \frac{\alpha }{\beta} \left(1-a + ae^{-\beta/a}-e^{-\beta}\right)\\
				& = \frac{\alpha }{\beta}\left(\left(1-e^{-\beta}\right) - a\left(1-e^{-\beta/a}\right)\right)\ge 0.
			\end{split}
		\end{equation*}
		
		
		\item By \citet[Proposition 3.9]{CT2003} a \Levy\ process with characteristic triplet $\left(A,\nu,\gamma\right)$ is of finite variation if and only if:
		\begin{equation*}
			A = 0 \quad \text{and} \quad \int_{|x|\le 1} |x| \nu\left(dx\right) <\infty.
		\end{equation*}
		$A=\sigma_a=0$ and the computation of the integral is straightforward:
		\begin{equation*}
			\begin{split}
				\int_{|x|\le 1} |x| \nu_{a}\left(dx\right)  & = \int_{0}^{1} \alpha t\left(e^{-\beta x} - e^{-\beta x/a}\right)dx \\ 
				&= \frac{\alpha }{\beta} \left(1-e^{-\beta} -a\left(1-e^{-\beta/a}\right)\right)<\infty
			\end{split}
		\end{equation*}
		By \citet[Proposition~3.10]{CT2003} since $\sigma_{a}=0$, $\nu_{a}\left(\left(-\infty,0\right]\right)=0$ and $b = \gamma - \int_{0}^{1}x\nu_{a}\left(x\right)\ge 0$ it follows that $Z_{a} $ is a subordinator.
		
		\item As a direct consequence of \citet[3.434]{gradshteyn2007} we have:
		\begin{equation*}
			\nu_{a}\left(\R\right) = \alpha \int_{-\infty}^{\infty} \frac{e^{-\beta x} - e^{-\beta x/a}}{x}\mathbbm{1}_{\left(0,\infty\right)}\left(x\right)dx = \log\left(\frac{1}{a}\right)< \infty,
		\end{equation*}
		hence, $\zapp$ has finite activity and is a compound Poisson process such that $\nu\left(x\right) = \lambda f\left(x\right)$ where $f(x)$ represents the \pdf\ of the jumps. Define $\Lambda  = \log\left(1/a\right)$, it follows that:
		\begin{equation*}
			\begin{split}
				\nu_{a}\left(x\right) & = \Lambda \alpha \cdot  \frac{1}{\Lambda x} \left(e^{-\beta x} - e^{-\beta/a x}\right) = \Lambda \alpha \cdot  \frac{1}{\Lambda x} \int_{1}^{1/a} -\beta x e^{-\beta x y}dy \\
				& = \Lambda \alpha \int_{1}^{1/a} -\frac{\beta}{\Lambda}e^{-\beta x y}dy  = \Lambda \alpha \int_{1}^{1/a} \frac{\beta}{\log a} e^{-\beta x y} dy \\
				& = \Lambda \alpha \int_{1}^{1/a} \frac{\beta y}{y \cdot \log a} e^{-\beta x y} dy \\
				& = \underbrace{ \vphantom{\int_{0}^{1}}\Lambda \alpha}_{ \lambda} \cdot \underbrace{\int_{1}^{1/a} \frac{1}{y \log a} f_{\mathcal{E}}\left(x| \mu  = \beta y\right)dy}_{f\left(x\right)},
			\end{split}
		\end{equation*}
		where $f_{\mathcal{E}}\left(x| \mu\right)$ is the \pdf\ of an exponential distribution with parameter $\mu>0$ and that concludes the proof.
		\end{enumerate}
\end{proof}

We remark that Proposition \ref{prop:chtripletteZa} \textit{(iii)} states that the distribution of the jump sizes can be seen as a mixture of an exponential law with stochastic rate given by $\beta Y$ where $Y$ is a \rv\ whose \pdf\ is given by $g_{Y}(y) = \frac{1}{y \log a}\mathbbm{1}_{\left[1,1/a\right]}(y)$. 
The cumulative distribution function of $Y$ is given by:
\begin{equation*}
	F_{Y}\left(x\right) = \frac{1}{\log a} \int_{1}^{x}\frac{1}{y}dy= \frac{\log x}{\log a},
\end{equation*} 
and it is then easy to verify that
\begin{equation*}
    Y \eqd a^U, \quad U\sim\unif([0, 1]),
\end{equation*} 
which simplifies the simulation of the skeleton of the process $\zapp$ as illustrated in Algorithm~\ref{algo:ZaSimulationCompoundPoisson}.


\begin{algorithm}
	\caption{Simulation of $Z_{a}\left(t\right)$.}
	\label{algo:ZaSimulationCompoundPoisson}
	\begin{algorithmic}[1]
		\State Simulate $n \sim \poiss\left(\alpha t \log\left(1/a\right)\right)$
		\State Simulate $n$ \iid\ \rv's $u_i\sim\unif([0,1])$ and set $y_i=a^{u_i}$
		\State Simulate $n$ \iid\ \rv's $J_{i}\sim\erl(\beta\,y_i)$.
		\State Set $\zapp(t) =  \sum_{i=0}^{n}J_{i}$.
	\end{algorithmic}
\end{algorithm}
Alternatively, as shown in \citet{SabinoPetroni2020GammaRelated} the skeleton of $\zapp$ can be simulated as a stochastic sum of independent exponentially distributed \rv's with parameter $\beta/a$, where the number of terms is given by $S(t) \sim \pol\left(\alpha t, 1-a\right)$ as summarized in Algorithm \ref{algo:ZaSimulationPolyasum}.
\begin{algorithm}
	\caption{Simulation of $Z_{a}\left(t\right)$.}
	\label{algo:ZaSimulationPolyasum}
	\begin{algorithmic}[1]
		\State Simulate $s \sim \pol\left(\alpha t, 1-a\right)$.
        \State Set $\zapp(t) \sim\erl_s(\beta/a)$.
	\end{algorithmic}
\end{algorithm}

\section{Variance Gamma++ process}
\label{sec:VG++process}
\par In Section \ref{sec:preliminaries} we have shown that $\zapp$ is a subordinator and hence  can be used to time change a Brownian motion. 
\begin{defn}
Consider a Brownian motion $W=\left\{W(t);t\ge0\right\}$, with drift $\theta \in \R$, diffusion $\sigma \in \R^+$ independent of $\zapp$. We call the process $X = \left\{X(t);t\ge0\right\}$  defined as
\begin{equation}
	X(t) = \theta \zapp(t) + \sigma W\left(\zapp\left(t\right)\right), \qquad t \ge 0
	\label{eqn:VG++}
\end{equation}
\VGpp\ process.
\end{defn}

In the following we detail its properties.
\begin{prop}
	\label{prop:chfVG++}
	For $u \in \R$, the \chf\ of $X$ at time $t$ is given by:
	\begin{equation}
		\phi_{X(t)}\left(u\right)= \phi_{\zapp(t)}\left(\theta u +iu^2 \frac{\sigma^2}{2}\right) =
		\left(\frac{\beta -i\left(\theta u + i u^2 \sigma^2 /2\right)a}{\beta - i\left(\theta u + iu^2 \sigma^2 /2\right)}\right)^{\alpha t}.
		\label{eqn:chfVG++}
	\end{equation}
\end{prop}
\begin{proof}
	Knowing that the \chf\ of the Gaussian distribution $\norm(\mu, \sigma)$ is given by:
	\begin{equation*}
		\phi\left(u\right) = e^{i\theta u -\frac{\sigma^2 u^2}{2}},
		\label{eqn:chfNormalvariable}
	\end{equation*}
	and from Equation \eqref{eqn:chfgammareminder} that the \chf\ of $\zapp(t)$ is
	\begin{equation}
		\phi_{\zapp}(u) = \exp\left\{ t \log\left(\frac{\beta -iua}{\beta - iu}\right)^{\alpha}\right\},
		\label{eqn:chfZaprocess}
	\end{equation}
	we have:
	\begin{equation*}
		\begin{split}
			\E\left[e^{iuX(t)}\right] & = \E\left[e^{iu\left(\theta \zapp\left(t\right) + \sigma W\left(Z_{a}\left(t\right)\right)\right)}\right] = \E\left[\E\left[\left.e^{iu\left(\theta \zapp(t) + \sigma W\left(Z_{a}(t)\right)\right)}\right|\zapp(t)\right]\right] \\
			& = \E\left[e^{iu\theta \zapp(t) - \frac{\sigma^2}{2}u^{2}\zapp(t)}\right] = \E\left[e^{i\left(u\theta + i\frac{\sigma^2}{2}u^2\right)\zapp(t)}\right] = \phi_{\zapp(t)}\left(u\theta + i\frac{\sigma^2}{2}u^2\right)\\
			& = \exp\left\{ \log\left(\frac{\beta -i\left(u\theta + i\frac{\sigma^2}{2}u^2\right)a}{\beta - i\left(u\theta + i\frac{\sigma^2}{2}u^2\right)}\right)^{\alpha t}\right\} 
			= \left(\frac{\beta -i\left(\theta u + i u^2 \sigma^2 /2\right)a}{\beta - i\left(\theta u + iu^2 \sigma^2 /2\right)}\right)^{\alpha t}.
		\end{split}
	\end{equation*}
	that concludes the proof.
\end{proof}


\begin{prop}
	\label{prop:VGppdifferenceofSubordinators}
	The \VGpp\ process can be written as difference of two independent processes $Z_{a_{p}}^{++} = \left\{Z_{a_{p}}^{++}(t);t\ge 0\right\}$ and $Z_{a_{n}}^{++} = \left\{Z_{a_{n}}^{++}(t);t\ge 0\right\}$ where $Z_{a_{p}}^{++}\left(t\right)\sim \Gamma^{++}\left(a_p, \alpha t,\beta_{p}\right)$ and $Z_{a_{n}}^{++}\left(t\right)\sim \Gamma^{++}\left(a_n, \alpha t,\beta_{n}\right)$.
\end{prop} 
\begin{proof}
Given the definition of the \chf\ of $X(t)$, it results
	\begin{equation*}
		\phi_{X(t)}(u) = \phi_{\zapp}\left(u\theta + \frac{iu^2 \sigma^2}{2}\right) = \frac{\left(\frac{1}{1 - \frac{i}{\beta}\left(u\theta + \frac{i u^2 \sigma^2}{2}\right)}\right)^{\alpha t}}{\left(\frac{1}{1 - \frac{ia}{\beta}\left(u\theta + \frac{i u^2 \sigma^2}{2}\right)}\right)^{\alpha t}} = \frac{A}{B}.
	\end{equation*}
	Consider the term $A$: 
	\begin{equation*}
		A = \left(\frac{1}{1 - \frac{i}{\beta}\left(u\theta + \frac{i u^2 \sigma^2}{2}\right)}\right)^{\alpha t} = 
		\left(\frac{1}{1-\frac{iu}{\beta_p}}\right)^{\alpha t}\left(\frac{1}{1+\frac{iu}{\beta_n}}\right)^{\alpha t},
	\end{equation*}
	and its denominator
	\begin{equation*}
		1 -iu\frac{\theta}{\beta} -i^2u^2\frac{\sigma^2}{2\beta} = 1 -iu\left(\frac{1}{\beta_{p}} - \frac{1}{\beta_n}\right) - iu^2\frac{1}{\beta_{p}\beta_{n}}.
	\end{equation*}
	It turns out then:
	\begin{equation*}
		\frac{\theta}{\beta}  = \frac{1}{\beta_{p}} - \frac{1}{\beta_{n}}, \qquad
		\frac{1}{\beta_p \beta_n}  = \frac{\sigma^2}{2\beta}.
	\end{equation*}
	Solving the previous system of equations with respect to $\beta_p$ and $\beta_n$ and taking only the positive solution we have that:
	\begin{equation*}
		\beta_n= \frac{\sqrt{\theta^2 + 2\sigma^2 \beta} +\theta}{\sigma^2}, \qquad \beta_p= \frac{ \sqrt{\theta^2 + 2\sigma^2 \beta} - \theta}{\sigma^2}.
	\end{equation*}
	Similarly, the term $B$ can be decomposed as:
	\begin{equation*}
		\tilde{\beta}_n= \frac{\sqrt{\theta^2 + 2\sigma^2 \beta/a} +\theta}{\sigma^2}, \qquad \tilde{\beta}_p= \frac{ \sqrt{\theta^2 + 2\sigma^2 \beta/a} - \theta}{\sigma^2}.
	\end{equation*}
	It follows that:
	\begin{equation}
		\phi_{X\left(t\right)} = \frac{\left(\dfrac{1}{1-iu/\beta_p}\right)^{\alpha t}\left(\dfrac{1}{1+iu/\beta_n}\right)^{\alpha t}}{\left(\dfrac{1}{1-iu/\tilde{\beta}_p}\right)^{\alpha t}\left(\dfrac{1}{1+iu/\tilde{\beta}_n}\right)^{\alpha t}} = \left(\frac{1-iu\left(\dfrac{\beta_p}{\tilde{\beta}_{p}}\right)/\beta_{p}}{1-\dfrac{iu}{\beta_{p}}}\right)^{\alpha t} \left(\frac{1+iu\left(\dfrac{\beta_n}{\tilde{\beta}_{n}}\right)/\beta_{n}}{1+\dfrac{iu}{\beta_{n}}}\right)^{\alpha t}.
		\label{eqn:betatildeexpression}
	\end{equation}
	Because  $0<\beta_{p}/\tilde{\beta}_{p}<1$ we can define $a_{p} = \beta_{p}/\tilde{\beta}_{p}$ and $a_{n} = \beta_{n}/\tilde{\beta}_{n}$ and we obtain: 
	\begin{equation*}
		\phi_{X(t)}(u) = \left(\frac{1- iua_{p}/\beta_{p}}{1-iu/\beta_{p}}\right)^{\alpha t}\left(\frac{1+ iua_{n}/\beta_{n}}{1+iu/\beta_{n}}\right)^{\alpha t}	
	\end{equation*}
	which is the \chf\ of the difference of two independent \rv's $Z_{a_{p}}^{++}(t) \sim \Gamma_{a_{p}}^{++}\left(\alpha t, \beta_{p}\right)$ and $Z_{a_{n}}^{++}(t) \sim \Gamma_{a_{n}}^{++}\left(\alpha t, \beta_{n}\right)$. Therefore the process $X$ can be expressed as difference of two independent subordinators $Z_{a_{p}}^{++} = \left\{Z_{a_{p}}^{++}(t); t\ge 0\right\}$ and $Z_{a_{n}}^{++} = \left\{Z_{a_{n}}^{++}(t); t\ge 0\right\}$.
\end{proof}

\begin{prop}
	The \Levy\ measure of the \VGpp\ process $X$ is given by:
	
	\begin{equation*}
		\begin{split}
			\nu\left(x\right) &= \left(\alpha x^{-1} e^{-x\beta_{p}} - \alpha x^{-1} e^{-x\beta_{p}/a_{p}}\right)\mathbbm{1}_{\left(0,\infty \right)}(x) \\
			& + \left(-\alpha x^{-1} e^{x \beta_{n}} + \alpha x^{-1} e^{x\beta_{n}/a_{n}}\right)\mathbbm{1}_{\left(-\infty,0 \right]}(x).
		\end{split}
	\end{equation*}
	The process $X$ is of finite activity and therefore of finite variation.
\end{prop}
\begin{proof}
The proof is a simple consequence of Proposition~\ref{prop:VGppdifferenceofSubordinators} and Proposition~\ref{prop:chtripletteZa}.
	
\end{proof}

We recall that the cumulant generating function $\psi_{Y}\left(u\right)$ and the cumulants of a \rv\ $Y$ with \chf\ $\phi_{Y}\left(u\right)$ are defined, respectively, as:
\begin{equation*}
	\psi_{Y}\left(0\right)=0, \quad \phi_{Y}\left(u\right)=e^{\psi_{Y}\left(u\right)},
\end{equation*}
\begin{equation*}
	c_{n}\left(X\right) = \frac{1}{i^{n}} \frac{\partial^{n} \psi_{X}}{\partial u^{n}}\left(0\right).
\end{equation*}
%

\begin{prop}
	\label{prop:cumulants}
	The first four cumulants  of the process $X$ at time $t\ge 0$ are given by:
	\begin{equation*}
	\begin{split}
		c_{1}\left(X(t)\right) & = \E\left[X(t)\right] = \alpha t \left(\frac{1}{\beta_{p}} - \frac{1}{\tilde{\beta}_{p}} - \frac{1}{\beta_{n}} + \frac{1}{\tilde{\beta}_{n}}\right), \\		c_{2}\left(X(t)\right) & = Var\left[X(t)\right] = \alpha t \left(\frac{1}{\beta_{p}^2} - \frac{1}{\tilde{\beta}_{p}^2} + \frac{1}{\beta_{n}^2} - \frac{1}{\tilde{\beta}_{n}^2}\right), \\
		c_{3}\left(X(t)\right) & = 2\alpha t \left(\frac{1}{\beta_{p}^3} - \frac{1}{\tilde{\beta}_{p}^3} - \frac{1}{\beta_{n}^3} + \frac{1}{\tilde{\beta}_{n}^3}\right), \\
		c_{4}\left(X(t)\right) & = 6\alpha t \left(\frac{1}{\beta_{p}^4} - \frac{1}{\tilde{\beta}_{p}^4} + \frac{1}{\beta_{n}^4} - \frac{1}{\tilde{\beta}_{n}^4}\right), \\
	\end{split}	
	\end{equation*}
where $\beta_{p}$, $\tilde{\beta}_{p}$, $\beta_{n}$, $\tilde{\beta}_{n}$ are defined in Proposition \ref{prop:VGppdifferenceofSubordinators}. 
\end{prop}

\begin{proof}
Using \citet[Proposition~13.3]{CT2003} and Proposition \ref{prop:characteristictripletarem}, it results that if the law of $Y$ is \sd\ the $n$-th cumulant of the \arem\  $Z_a$ is:
\begin{equation}
	\begin{split}
	c_{n}\left(Z_a\right)= t \int_{-\infty}^{\infty} x^{n}\nu_{a}\left(x\right)\,dx = \left(1-a^{n}\right)c_{n}\left(Y\right),
	\end{split}
\label{eqn:procumulants1}
\end{equation} 
where $\nu_{a}\left(x\right)$ is the \Levy\ measure of $Z_{a}$.\\
Moreover, it is easy to prove that for two independent \rv's $X$ and $Y$ with finite cumulants of order $n$, taking $U = X-Y$, it holds:
\begin{equation}
	c_{n}(U) = c_{n}(X) + (-1)^{n}c_{n}(Y).
	\label{eqn:procumulants2}
\end{equation}
Combining \eqref{eqn:procumulants1} and \eqref{eqn:procumulants2} and the fact that from Proposition \ref{prop:VGppdifferenceofSubordinators} the \VGpp\ process can be written as the difference of two independent subordinators $Z_{a_{p}}^{++}$ and $Z_{a_{n}}^{++}$ it results
\begin{equation*}
	c_{n}\left(X(t)\right) = \left(1-a_{p}^{n}\right) c_{n}\left(G_{1}(t)\right) + (-1)^{n} \left(1-a_{n}^{n}\right) c_{n}\left(G_{2}(t)\right),
\end{equation*}
where $G_{1}=\left\{G_{1}\left(t\right);t\ge 0\right\}$ and  $G_{2}=\left\{G_{2}\left(t\right);t\ge 0\right\}$ are Gamma processes with parameters $\left(\alpha,\beta_{p}\right)$ and  $\left(\alpha,\beta_{n}\right)$ respectively. The proof is simply concluded recalling the expression of the cumulants of the gamma laws $\Gamma(\alpha t, \beta_p)$ and $\Gamma(\alpha t, \beta_n)$, respectively:
\begin{align*}
	c_{n}\left(G_{1}(t)\right) & = (n-1)!\frac{\alpha t}{\beta_{p}^{n}},\\
	c_{n}\left(G_{2}(t)\right) & = (n-1)!\frac{\alpha t}{\beta_{n}^{n}}.
\end{align*}
\end{proof}

\begin{prop}
	\label{prop:pdfofVgppProcess}
	The \pdf\ of the \VGpp\ process $X = \left\{X(t); t\ge 0\right\}$ at $t\ge 0$ is given by:
	\begin{equation}
		f_{X(t)}(x) = a^{\alpha t} \delta_{0}\left(x\right) + \sum_{k\ge1}\binom{\alpha t + k -1}{k}a^{\alpha t}\left(1-a\right)^{k} f^{VG}_{k,\beta/a}\left(x\right).
		\label{eqn:VG++Density}
	\end{equation}  
	where $\delta_{0}(x)$ is the Dirac function and $f^{VG}_{k,\beta/a}\left(x\right)$ is the \pdf\ of a Variance Gamma law with parameters $k\in \mathbb{N}$ and $\beta/a$ which is given by:
	\begin{equation*}
		\begin{split}
			f_{k,\beta/a}^{VG}\left(x\right) =
			 K_{k-\frac{1}{2}}\left(|x| \frac{\sqrt{2 \sigma^2 \beta/a + \theta^2}}{\sigma^2}\right) 
			\frac{\exp\left(\theta x/\sigma^2\right)}{\sqrt{2 \pi \sigma^2}} \frac{\left(\beta/a\right)^{k}}{\Gamma\left(k\right)} \left(2 \sigma^2 \beta + \theta^2 \right)^{\frac{1}{4} - \frac{k}{2}} 2 |x| ^{k -\frac{1}{2}}.
		\end{split}
	\end{equation*} 
\end{prop}

\begin{proof}
From Equation \eqref{eqn:chfVG++} we have that:
		
		\begin{equation}
			\begin{split}
				\phi_{X(t)}\left(u\right) &= 
				\left(\frac{\beta -i\left(\theta u + i u^2 \sigma^2 /2\right)a}{\beta - i\left(\theta u + iu^2 \sigma^2 /2\right)}\right)^{\alpha t} = \left(\frac{a}{1-\left(1-a\right)\frac{\beta}{\beta-ia\left(\theta u + iu^2\sigma^2/2\right)}}\right)^{\alpha t}\\
				&=\sum_{k=0}^{\infty}\binom{\alpha t + k -1}{k} a^{\alpha t} \left(1-a\right)^{k}\left(\frac{\beta}{\beta - ia\left(\theta u +iu^2 \sigma^2/2\right)}\right)^{k}\\
				& = a^{\alpha t} + \sum_{k\ge 1}\binom{\alpha t + k -1}{k} a^{\alpha t} \left(1-a\right)^{k}\left(\frac{\beta}{\beta - ia\left(\theta u +iu^2 \sigma^2/2\right)}\right)^{k}.
			\end{split} 
		\label{eqn:chfVgppProof}
		\end{equation}
		One can notice that $X\left(t\right)$ is a mixture of Variance Gamma \rv's, where the weights are given by a Polya distribution plus a degenerate distribution at $x=0$. By taking the inverse Fourier transform of \eqref{eqn:chfVgppProof} we get the \pdf\ in \eqref{eqn:VG++Density}.
		
	
\end{proof}

\begin{rem}
	\label{rem:simplifiedBessel}
	For $n\in\N$ the modified Bessel function of the second kind $K_{n+\frac{1}{2}}\left(x\right)$ can be written in terms of elementary functions (see \citet[pag. 443]{abramowitzstegun1964}):
	\begin{equation*}
		\sqrt{\frac{\pi}{2x}}K_{n+\frac{1}{2}}\left(x\right)= \left(\frac{\pi}{2x}\right)e^{-x}\sum_{k=0}^{n}\frac{\left(n+k\right)!}{k!\Gamma\left(n-k+1\right)}\left(2x\right)^{-k}.
	\end{equation*}
This fact is instrumental to obtain an efficient formula for the pricing of a European call option when the evolution of the market is modelled by a Variance Gamma process with $\frac{t}{\nu}\in\N$ and, as we shall show, by a \VGpp\ process as well.
\end{rem}

\vspace{0.3cm}
\begin{prop}
	Consider the \VGpp\ process $X$ and let $S$ be a Polya process such that $S(t)\sim \pol\left(\alpha t,1-a\right)$. In addition let $\left(I_{k}\right)_{k\ge 1}$ and $\left(J_{k}\right)_{k\ge 1}$ be two independent sequences of \iid\ \rv's, with $I_{k}\sim \erl\left(\tilde{\beta}_{p}\right)$, $J_{k}\sim \erl\left(\tilde{\beta}_{n}\right)$ where $\tilde{\beta}_{n}$ and $\tilde{\beta}_{p}$ are defined  in Equation \eqref{eqn:betatildeexpression}. Finally take $\delta_{k} = I_{k} - J_{k}$ and define the process $C = \left\{C(t);t \ge 0\right\}$ as:
	\begin{equation*}
		C(t) = \sum_{k=0}^{S(t)} \delta_{k},\quad C(t)=0 \text{ when } S(t)=0. 
	\end{equation*} 
	Then:
	\begin{equation*}
		X(t)\stackrel{d}{=} C(t),\; t>0.
	\end{equation*}
\end{prop}

\begin{proof}
    	First we prove that the \VGpp\ process at time $t$ can be written as a Polya sum of independent \rv's. For $u \in \R$, consider the \chf\ $\phi_{X(t)}\left(u\right)$ at time $t$ of the \VGpp\ process given in \eqref{eqn:chfVG++} and define $g(u) = i\left(\theta u + i u^2 \sigma^2 /2\right)$. We have:
	\begin{equation*}
		\begin{split}
		\phi_{X(t)}\left(u\right) &= \left(\frac{1}{\frac{\beta-g(u)}{\beta -ag(u)}}\right)^{\alpha t} = \left(\frac{a}{\frac{a\beta + \beta - \beta ag(u)}{\beta - ag(u)}}\right)^{\alpha t} = \left(\frac{a}{1-\left(1-a\right) \frac{\beta}{\beta - ag(u)}}\right)^{\alpha t} \\
		& \stackrel{a=1-p}{=} \left(\frac{1-p}{1-p\frac{1}{1-\frac{a}{\beta}g(u)}}\right)^{\alpha t} = \left(\frac{1-p}{1-p\varphi(u)}\right)^{\alpha t},
		\end{split} 	
	\end{equation*}
where:
\begin{equation*}
	\varphi(u) = \frac{1}{1-\frac{a}{\beta}g(u)} = \frac{\beta/a}{\beta/a -iu\theta + u^{2}\sigma^2/2}.
\end{equation*}
    Therefore, $X(t)$  can be represented as a Polya sum of independent \rv's whose \chf\ is given by $\varphi(u)$. We can write:
\begin{equation*}
	\varphi(u) = \frac{1}{1- \frac{iua\theta}{\beta} - \frac{i^2 u^{2}\sigma^2}{2}}
\end{equation*}
and the denominator can be decomposed as:
\begin{equation*}
	1-\frac{iua\theta}{\beta} - \frac{i^2 u^{2}\sigma^2}{2} = \left(1-\frac{iu}{\tilde{\beta}_{p}}\right)\left(1+\frac{iu}{\tilde{\beta}_{n}}\right) = 1-iu\left(\frac{1}{\tilde{\beta}_{p}}-\frac{1}{\tilde{\beta}_{n}}\right) -i^2 u^2 \frac{1}{\tilde{\beta}_{p}\tilde{\beta}_{n}}.
\end{equation*} 
Taking:
\begin{equation*}
	\frac{1}{\tilde{\beta}_{p}}-\frac{1}{\tilde{\beta}_{n}} = \frac{a\theta}{\beta}, \qquad \frac{1}{\tilde{\beta}_{p}\tilde{\beta}_{n}} = \frac{2\beta}{a \sigma^2},
\end{equation*}
solving with respect to $\tilde{\beta}_{n}$ and $\tilde{\beta}_{p}$ and considering only positive solutions we have:
\begin{equation*}
	\tilde{\beta}_{p} = \frac{\sqrt{\theta^2+2\sigma^2 - \beta/a}-\theta^2}{\sigma^2}, \quad 
	\tilde{\beta}_{n} = \frac{\sqrt{\theta^2+2\sigma^2 - \beta/a}+\theta^2}{\sigma^2}.
\end{equation*}
Finally, $\varphi(u)$ can be written as:
\begin{equation*}
	\varphi(u) = \frac{1}{1-\frac{iu}{\tilde{\beta}_{p}}} \cdot \frac{1}{1+\frac{iu}{\tilde{\beta}_{n}}},
\end{equation*}
which is the \chf\ of the difference of two independent exponentially distributed \rv's with parameters  $\tilde{\beta}_{p}$ and $\tilde{\beta}_{n}$ respectively. \\ By computing the \chf\ of $C\left(t\right)$ it is easy to check that:
\begin{equation*}
	\phi_{C(t)}(u) = \phi_{X(t)}(u),
\end{equation*}
that means that $X(t)\stackrel{d}{=}C(t)$ which concludes the proof.
\end{proof}
Finally, Table \ref{tbl:fullcharacterization} summarizes the properties of the processes $\zapp$ and \VGpp.

\begin{table}
	\centering
	\scriptsize
	\begin{tabular}{c|c|c}
		\toprule
		Model name & $\zapp$ process & \VGpp\ process $X$ \\
		\midrule
		Model type & \makecell{Finite variation \\ Finite activity \\ Subordinator} & \makecell{Finite variation \\ Finite activity} \\
		\midrule
		Parameters & \makecell{$\alpha>0$ shape, $\beta>0$ rate and \\ $a\in\left(0,1\right)$ \sd} & \makecell{$\alpha,\beta,a$ + $\theta$ drift  and $\sigma$ diffusion \\ of the Brownian motion} \\
		\midrule
		\Levy\ measure & $\nu_{a}\left(x\right) = \frac{\alpha}{x}\left(e^{-\beta x} - e^{-\beta x/a}\right)\mathbbm{1}_{\left(0,\infty\right)}\left(x\right)$ & $\begin{aligned}\nu\left(x\right) &= \left(\alpha x^{-1} e^{-x\beta_{p}} - \alpha x^{-1} e^{-x\beta_{p}/a_{p}}\right)\mathbbm{1}_{\left(0,\infty \right)}(x) \\
		& + \left(-\alpha x^{-1} e^{x \beta_{n}} + \alpha x^{-1} e^{x\beta_{n}/a_{n}}\right)\mathbbm{1}_{\left(-\infty,0 \right]}(x)\end{aligned}$ \\
		\midrule
		\makecell{\chf} & 
			$\phi_{Z_{a}(t)}\left(u\right) = \left(\frac{\beta-iua}{\beta -iu}\right)^{\alpha t}$ & $\phi_{X(t)}\left(u\right) =
			\left(\frac{\beta -i\left(\theta u + i u^2 \sigma^2 /2\right)a}{\beta - i\left(\theta u + iu^2 \sigma^2 /2\right)}\right)^{\alpha t}$ \\
		\midrule
		\makecell{\pdf} & 
			\makecell{$ \begin{aligned}
				 g_{a}\left(x\right) & = a^{\alpha}\delta_{0}\left(x\right) \\
				& + \sum_{n \ge 1} \binom{\alpha + n -1}{n} a^{\alpha} \left(1-a\right)^{n} \\ & \cdot f_{n,\beta/a}\left(x\right) \mathbbm{1}_{\left(0,\infty\right)}\left(x\right)dx
			\end{aligned}$ \\ \\
		where $f_{n,\beta/a}\left(x\right)$ is
		the density of\\  an Erlang distribution.} 
		 & \makecell{$\begin{aligned}f_{X(t)}(x) & = a^{\alpha t} \delta_{0}\left(x\right) \\
		 	& + \sum_{n\ge1}\binom{\alpha t + n -1}{n}a^{\alpha t}\left(1-a\right)^{n} \\
		 	& \cdot
		 	  f^{VG}_{n,\beta/a}\left(x\right)dx \end{aligned}$ \\ \\ where $f^{VG}_{n,\beta/a}\left(x\right)$ is the density of \\ a Variance Gamma distribution.} \\
		\midrule
		 \makecell{Cumulants} & \makecell{$\begin{aligned}
		 		c_{1}\left(\zapp(t)\right) & = \alpha t\frac{1-a}{\beta}, \\		c_{2}\left(\zapp(t)\right) & = \alpha t\frac{1-a^2}{\beta^2}, \\
		 		c_{3}\left(\zapp(t)\right) & = 2\alpha t\frac{1-a^3}{\beta^3}, \\
		 		c_{4}\left(\zapp(t)\right) & = 6\alpha t\frac{1-a^4}{\beta^4}. \\
		 	\end{aligned}$\\
 			} & \makecell{$\begin{aligned}
		 	c_{1}\left(X(t)\right) & =  \alpha t \left(\frac{1}{\beta_{p}} - \frac{1}{\tilde{\beta}_{p}} - \frac{1}{\beta_{n}} + \frac{1}{\tilde{\beta}_{n}}\right), \\		c_{2}\left(X(t)\right) & = \alpha t \left(\frac{1}{\beta_{p}^2} - \frac{1}{\tilde{\beta}_{p}^2} + \frac{1}{\beta_{n}^2} - \frac{1}{\tilde{\beta}_{n}^2}\right), \\
		 	c_{3}\left(X(t)\right) & = 2\alpha t \left(\frac{1}{\beta_{p}^3} - \frac{1}{\tilde{\beta}_{p}^3} - \frac{1}{\beta_{n}^3} + \frac{1}{\tilde{\beta}_{n}^3}\right), \\
		 	c_{4}\left(X(t)\right) & = 6\alpha t \left(\frac{1}{\beta_{p}^4} - \frac{1}{\tilde{\beta}_{p}^4} + \frac{1}{\beta_{n}^4} - \frac{1}{\tilde{\beta}_{n}^4}\right). \\
		 \end{aligned}$\\
	 	\vspace{0.5cm}
	 	\\ with $\tilde{\beta}_{n},\tilde{\beta}_{p},\beta_{n},\beta_{p}$ as in Proposition \ref{prop:VGppdifferenceofSubordinators}.}  \\
		\bottomrule  
		\end{tabular}
	\caption{Characterization of $\zapp$ and of the \VGpp\ process.}
	\label{tbl:fullcharacterization}
\end{table}

\subsection{An option pricing formula under the \VGpp\ model}
\label{sec:VG++closedformula}
Following \citet{CT2003}, we model the evolution of a risky asset by the process $F = \left\{F(t);t \ge 0 \right\}$ defined as
\begin{equation}\label{eq:market:ct}
	F\left(t\right) = F\left(0\right)e^{rt + \omega t + \theta \zapp(t) + \sigma W\left(\zapp(t)\right)} = F\left(0\right)e^{rt + \omega t + X(t)},
\end{equation}
where:
\begin{equation*}
	\omega  = \log\left(\frac{\beta -\left(\theta +\sigma^2/2\right)}{\beta - a\left(\theta + \sigma^2/2\right)}\right)^{\alpha},
\end{equation*}
to have non-arbitrage conditions. 

The following proposition provides a closed formula for the price of a European call option.

\begin{prop}\label{prop:call}
	Consider the market model of Equation \eqref{eq:market:ct} where $X$ is a \VGpp\ process, the price at time $0$ of a European call option with strike price $K$ and maturity $T$  is given by:
	\begin{equation}
		C\left(0,K\right) = C\left(0\right)a^{\alpha T} + \sum_{n\ge 1}\binom{\alpha T +n -1}{n}\left(1-a^{n}\right)a^{\alpha T} C_{n,\beta/a}^{VG}(0,K),
		\label{eqn:optionpricingFormula}
	\end{equation}
    where
    \begin{equation*}
    	C\left(0\right) = \max\left(F(0)e^{\omega T} - e^{-rT}K,0\right)
    \end{equation*}
	and $C_{n,\beta/a}^{VG}(0,K)$ is the price of a call option with strike $K$ and maturity $T$ under the Variance Gamma model with parameters $n$ and $\beta/a$.
\end{prop}

\begin{proof}
	Consider $X(T)=\theta \zapp(T) + \sigma W(\zapp(T))$ whose \pdf\ $f_{X(T)}(x)$ is given by Equation \eqref{eqn:VG++Density}. The value of the call option at $t=0$ is the discounted expected value under the risk-neutral measure:
	\begin{equation*}
		\begin{split}
			C\left(0,T\right) & = e^{-rT}\E\left[\left(F(T)-K\right)^{+}\right] = e^{-rT}\int_{-\infty}^{\infty} \left(F(0)e^{rT+\omega T + x}-K\right)^{+}f_{X(T)}(x)dx \\
			& = e^{-rt}\int_{-\infty}^{\infty}\left(FS(0)e^{rT+\omega T + x}-K\right)^{+}a^{\alpha T}\delta_{0}\left(x\right) \\
			& + e^{-rT}\int_{-\infty}^{\infty} \left(F(0)e^{rT+\omega T x}-K\right)^{+} \cdot \left(\sum_{n \ge 1} \binom{\alpha T +n -1}{n} a^{\alpha T}\left(1-a\right)^{n}f_{n,\beta/a}^{VG}\left(x\right)\right)dx\\
			& = \underbrace{a^{\alpha T}\left(F(0)e^{\omega T}-e^{-rT}K\right)^{+}}_{C(0)} \\
			& + \sum_{n \ge 1} \binom{\alpha T+n-1}{n}a^{\alpha T}\left(1-a\right)^{k} \underbrace{\int_{-\infty}^{\infty} \left(F(0)e^{rT+\omega T + x}-K\right)^{+}f_{k,\beta/a}(x)dx}_{C_{n,\beta/a}^{VG}\left(0,T\right)}
		\end{split}
	\end{equation*}
where in the last step we used the monotone convergence theorem to interchange the order of the integral and the summation.
\end{proof}

\begin{rem}
	The option price in Equation \eqref{eqn:optionpricingFormula} can be computed in a very efficient way using the results about \emph{EPT}-distributions discussed in \citet{SextonHanzon12} and summarized in Appendix \ref{sec:VarianceErlangOptionFormula}. Indeed, when the shape parameter $n\in \N$, the computation of $C_{n,\beta/a}^{VG}\left(0,T\right)$ is easier than when it is a real number. This fact directly stems from what we observed in Remark \ref{rem:simplifiedBessel}, namely that the Bessel function $K_{n}\left(x\right)$ can be written as a sum  of elementary functions when $n\in \N$.  The advantage is that one  does not need to compute any integral when we evaluate $C_{n,\beta/a}^{VG}\left(0,T\right)$ because this term can be simply obtained as matrix multiplications which are faster than numerical integration. 
	
	Table \ref{tbl:speed} shows the comparison of the computational times required to price a call option  when the shape parameter is either an integer or a positive real number  using  \textit{MATLAB} on a PC with an Intel Core i5-10210U 2.11 GHz processor. Apparently, the computation taking an integer shape parameter is  $10^{4}$ times faster.
	
	\begin{table}
		\centering
		\small
		\begin{tabular}{cc}
			\toprule
			Shape parameter domain & Computational time (s) \\
			\midrule
			$\N$ &  $7.61 \cdot 10^{-7}$  \\
			$\R$ &  $3.02 \cdot 10^{-3}$  \\
			\bottomrule
		\end{tabular}
		\caption{Computational time to price a European option if the shape parameter is a real or a natural number.}
		\label{tbl:speed}
	\end{table}
	
\end{rem}


\subsection{\VGpp\ backward simulation}
\label{sec:backwardsim}
So far, we have presented algorithms for the simulation of the trajectories of the \VGpp\ process forward in time over a given time grid $t_0, t_1, \dots, t_d$. On the other hand, we are not restricted to generate the random points of the trajectory in sequence, the only strict requirement is to generate points with the correct transition density.

In this section we illustrate how to simulate the \VGpp\ process  backward in time taking advantage of the notion of \emph{\Levy\ random bridges} (see \citet{Hoyle2010} for details), which are stochastic processes  pinned to a  fixed point at a fixed future time. 
Applications of \Levy\ bridge-based techniques in finance are for instance, the pricing with Monte Carlo (MC) methods of barrier options with continuous monitoring to avoid the bias arising by the use of the Euler discretization scheme, or the combination with Quasi-Monte Carlo methods (see for instance \citet{CMO1997} and \citet{Glass2004}).  

\Levy\ bridges naturally lead to the construction of \emph{backward simulations} as described in  \citet{PellegrinoSabino15}, \citet{HuZhou17} and \citet{Sabino20}. In principle, the computational cost of backward and forward strategies is the same, however numerical analysis showed that in most cases the forward construction is the faster solution (see \citet{Sabino20}). 

On the other hand, the path generation is only one component
of the overall pricing of derivative contracts with MC simulations. When the pricing
of contracts with complex American optionality is based on the Least Squares
Monte Carlo (LSMC) approach introduced by \citet{LSW01}, what matters in the
stochastic dynamic programming is the comparison between the intrinsic value and
the continuation value at a given time step $t$. If, for instance, we consider a $F$-factor
market model and we want to price an American option with LSMC, each step of
the Bellman backward recurrence requires to know the simulated prices or indices at two consecutive times $t$ and $t+\Delta$, nothing else. To this end, the forward generation requires storing $d\times N\times F$ numbers
where $d$ is the number of time steps and $N$ is the number of simulations, whereas
the backward solution requires storing a far lower number, $2 \times N \times F$. The forward construction may become computationally unfeasible for contracts with long maturities in contrast, although sometimes slower,  the backward construction is more reliable because one could generate a far higher number of trajectories that is often necessary for the computation of the Greek letters. 

\par In order to conceive a backward simulation scheme for the \VGpp\ process we start showing how to simulate the process $\zapp$ backward in time. Indeed, the backward simulation of the \VGpp\ will then consist of applying the well-known backward simulation of a Brownian motion on the stochastic grid generated by $\zapp$.




\begin{prop}[Polya Bridge]
	Consider a process $S = \left\{S(t); t \ge 0\right\}$ such that $S\left(0\right)=0$ a.s. and $S\left(t\right) \sim \pol\left(\alpha t, 1-a\right)$. 
	For $0<t\le T$, define the \rv\ $S_{tT}^{(k)}$, $k\in\N$ with probability mass function:
	\begin{equation*}
		\mathbb{P}\left(S_{tT}^{(k)} = j\right) \coloneqq \mathbb{P}\left(\left.S\left(t\right)= j\right| S\left(T\right)=k\right).
	\end{equation*}
	It results:
	
	\begin{equation*}
		\mathbb{P}\left(S_{tT}^{(k)} = j\right) = \binom{k}{j}\frac{\Beta\left(\alpha t + j, \alpha\left(T-t\right)+k-j\right)}{\Beta\left(\alpha t, \alpha \left(T-t\right)\right)}
	\end{equation*}
	namely, $S_{tT}^{\left(k\right)}$ is distributed according to  a beta-binomial law  
	$\mathcal{B}\left(\alpha t,\alpha\left(T-t\right),k\right)$ where $\Beta\left(\alpha,\beta\right)$ denotes the Beta function (see \citet{abramowitzstegun1964}).
\end{prop}\label{prop:polya:bridge}
\begin{proof}
	Knowing that $S$ has independent and stationary increments, the proof is verified as follows:
	\begin{equation*}
		\small
		\begin{split}
			\mathbb{P}\left(S_{tT}^{k}=j\right) & = \frac{P\left(S(t)=j,S(T)=k\right)}{\mathbb{P}\left(S(T)=k\right)}	 = \frac{\mathbb{P}\left(S(t)=j\right) \mathbb{P}\left(S(T-t)=k-j\right)}{\mathbb{P}\left(S(T)=k\right)} \\
			& = \frac{\binom{\alpha t +j -1}{j}\binom{\alpha\left(T-t\right)+k-j-1}{k-j}}{\binom{\alpha T+k-1}{k}} = \frac{\frac{(\alpha t + j -1)(\alpha t + j -2)\dots (\alpha t)}{j!} \cdot \frac{\left(\alpha(T-t)+k-j-1\right)\left(\alpha(T-t)+k-j-2\right)\dots \alpha(T-t)}{(k-j)!}}{\frac{(\alpha T+k-1)(\alpha T+k-2)\dots \alpha T}{k!}} \\
			&= \binom{k}{j} \frac{(\alpha t + j -1)(\alpha t + j -2)\dots \alpha t\cdot \left(\alpha(T-t)+k-j-1\right)\left(\alpha(T-t)+k-j-2\right)\dots \alpha(T-t)}{(\alpha T+k-1)(\alpha T+k-2)\dots \alpha T} \\
			& = \binom{k}{j} \frac{\Gamma\left(\alpha t+ j\right)}{\Gamma\left(\alpha t\right)} \frac{\Gamma\left(\alpha(T-t)+k-j\right)}{\Gamma\left(\alpha (T-t)\right)}\frac{\Gamma\left(\alpha T\right)}{\Gamma\left(\alpha T+k\right)} \\
			& = \binom{k}{j}\frac{\Beta\left(\alpha t +j,\alpha(T-t)+k-j\right)}{\Beta\left(\alpha t,\alpha(T-t)\right)},
		\end{split}
	\end{equation*}
	where we used the relations:
	\begin{equation*}
		\left(\alpha t +j -1\right)\left(\alpha t +j-2\right)\dots \alpha t =  \frac{\Gamma\left(\alpha t+j\right)}{\Gamma\left(\alpha t\right)},\qquad \frac{\Gamma\left(x\right)\Gamma\left(y\right)}{\Gamma\left(x+y\right)} = \Beta\left(x,y\right).
	\end{equation*}
\end{proof}

Based on Proposition~\ref{prop:polya:bridge} we can show that the process $\zapp$ is a gamma process $G$ subordinated by a Polya process $S$. This simple fact provides us with an easy way to simulate the process $\zapp$.

\begin{prop}
	\label{prop:ZaasGS}
	Consider a gamma process $G=\left\{G(t); t \ge 0\right\}$, such that $G\left(t\right) \sim \Gamma\left(t,\beta/a\right)$, $\beta>0$, $a \in \left(0,1\right)$, and  a Polya process $S = \left\{S(t); t\ge 0\right\}$ such that $S\left(t\right) \sim \pol\left(\alpha t,1-a\right)$. Define the process $Y = \left\{Y(t);t\ge0\right\}$ as:
	\begin{equation*}
		Y(t) = G\left(S(t)\right), \qquad t\ge0.
	\end{equation*}
	It results:
	\begin{equation*}
		\zapp\left(t\right) \stackrel{d}{=} Y\left(t\right),\qquad t\ge 0,
	\end{equation*}
where $\zapp$ is the \Levy\ process associated to the \arem\ of a gamma law with parameters $\alpha$ and $\beta$, as defined in \eqref{eqn:ZaProcess}.
\end{prop}

\begin{proof}
	We compute the \chf\ of $Y\left(t\right)$ for $u \in \R$. 
	
	\begin{equation}
		\begin{split}
			\E\left[e^{iuY\left(t\right)}\right] & =\E\left[\E \left[\left.e^{iuG\left(S(t)\right)}\right| S(t)\right]\right]
			 = \E\left[\left(\frac{\beta}{\beta- iua}\right)^{S(t)}\right] \\ &= \E\left[\left(\frac{\beta/a}{\beta/a -iu}\right)^{S\left(t\right)}\right].
		\end{split}
		\label{eqn:chfprocessZa}
	\end{equation}
	From Proposition~\ref{prop:ZaRapresentation} we have that 
	\begin{equation*}
		\zapp(t) = \sum_{n=0}^{S(t)}E_{n},
	\end{equation*}
	where $E_{n}$ are \iid\ \rv's with exponential law with parameter $\beta/a$. The \chf\ of $\zapp(t)$ is given by:
	
	\begin{equation*}
		\begin{split}
			\E\left[e^{iu\zapp(t)}\right] & = \E\left[e^{iu \sum_{n=0}^{S(t)}E_{n}}\right] = \E\left[\E\left[\left.e^{iu \sum_{n=0}^{S(t)}E_{n}}\right|S(t)\right]\right] \\
			& = \E\left[\prod_{n=0}^{S(t)}\E\left[e^{iuE_{1}}\right]\right] =  \E\left[\prod_{n=0}^{S(t)} \frac{\beta/a}{\beta/a - iu}\right] = \E\left[\left(\frac{\beta/a}{\beta/a - iu}\right)^{S(t)}\right]
		\end{split}
	\end{equation*}
	which is the same as Equation \eqref{eqn:chfprocessZa}, therefore we can conclude that $Z_{a}(t) \stackrel{d}{=} Y(t)$.
\end{proof}

Proposition \ref{prop:ZaasGS} illustrates how to simulate the process $\zapp$ backward in time: first, one simulates Polya process $S$ backward in time, and second one simulates the gamma process $G$ backward in time  on the stochastic time grid generated by $S$ (see \citet{Sabino20} for the backward simulation of a gamma process).
\par Assume, indeed, that given $\zapp(0)=0$ the value of the process $\zapp$ at time $T$ is equal to $z_{T}$, then $\zapp(t)$, $t\in (0,T)$  can be simulated by generating  the Polya bridge at time $t$ in the first step and the gamma bridge at a random time $S(t) \in \left(0,S(T)\right)$ in the second step. This procedure is summarized in Algorithm \ref{algo:ZaSimulation}.
\begin{algorithm}
	\caption{Backward simulation of $Z_{a}$.}
	\label{algo:ZaSimulation}
	\begin{algorithmic}[1]
		\State Generate  $s_{T}\sim \pol\left(\alpha T, 1-a\right)$.
		\State Generate $z_T \sim \Gamma\left(s_{T},b/a\right)$ and set $\zapp\left(T\right) = z_T$.
		\State Consider $t\in \left(0,T\right)$ and $p \sim Beta\left(\alpha t, \alpha \left(T-t\right)\right)$.
		\State Simulate $s_{t} \sim Bin \left(s_{T},p\right)$.
		\State Simulate $\beta \sim Beta\left(s_{t},s_{T}-s_{t}\right)$.
		\State Set $\zapp\left(t\right) = z_T \, \beta$.
	\end{algorithmic}
\end{algorithm}

In a similar way, the backward simulation of the \VGpp\ process can be accomplished implementing the backward simulation of the Brownian motion oven a random grid given by the backward simulation of $\zapp$ as illustrated in Algorithm  \ref{algo:VG++BackwardSimulation}.

\begin{algorithm}
	\caption{Backward simulation of $X$.}
	\label{algo:VG++BackwardSimulation}
	\begin{algorithmic}[1]
		\State Set $X(0)=0$ and $\zapp(0)=0$.
		\State Simulate $\zapp\left(T\right)$ and $\zapp\left(t\right)$ using Algorithm \ref{algo:ZaSimulation}.
		\State Simulate $x_{T}\sim \norm\left(\theta \zapp(T),\sigma^{2} \zapp(T) \right)$.
		\State Simulate $x_{t} \sim \norm\left(x_{T}\frac{\zapp(t)}{Z_{a}(T)},\frac{Z_{a}(t)\left(\zapp(T)-Z_{a}(t)\right)}{\zapp(T)}\sigma^{2}\right)$.
		\State Set $X(t) = x_{t}$.
	\end{algorithmic}
\end{algorithm}

%

Table \ref{tbl:momentcomparison} compares the theoretical mean, variance, skewness and kurtosis of $X$ at time $T=1$ with the ones obtained by numerical forward and backward simulations, where we used the following set of parameters: $\theta=1.025$, $\sigma=0.2$, $\alpha=5$, $\beta=15$, $a=0.7$, and $10^{6}$ simulations.

	\begin{table}
	\centering
	\small
	\begin{tabular}{cccc}
		\toprule
		Moment & $T$ & $F$ & $B$\\
		\midrule
		$\E(X)$ &  $0.10250$ & $0.10234$ & $0.10234$   \\
		$Var(X)$ &  $0.01591$ & $0.01584$ & $0.01582$   \\
		$s(X)$ &  $1.73973$ & $1.73637$ & $1.73569$   \\
		$k(X)$ &  $7.11923$ & $7.12693$ & $7.09786$   \\
		\bottomrule
	\end{tabular}
	\caption{Comparison of theoretical moments $(T)$ of the \VGpp\ process with the numerical ones obtained by forward $(F)$ and backward $(B)$ simulations.}
	\label{tbl:momentcomparison}
\end{table}
\section{Financial applications}
\label{sec:financialapplications}
In this section we show concrete applications of the \VGpp\ model to energy markets. First, we price European call options using three different approaches: the closed formula of Proposition \ref{prop:call}, Monte Carlo (MC) simulations, and the FFT method of \citet{Carr99}. \\
Secondly, we calibrate the \VGpp\ model on historical data focusing on power future market quotations adopting the \textit{Maximum Likelihood Estimator} (\MLE)  approach. Finally, we fit the model on quoted vanilla contracts using the standard Non-Linear-Least-Squares (\NLLS) technique and then we price non standard derivatives with backward simulations.

\subsection{Option pricing methods}
\label{subsec:optionpricingmethods}
In this subsection we compare the following three different methods for vanilla options pricing:
\begin{itemize}
    \item The closed formula derived in Section \ref{sec:VG++closedformula}.
    \item The MC method relying upon the Algorithm \ref{algo:ZaSimulationPolyasum} to simulate the process $\zapp$.
    \item The FFT method of \citet{Carr99} based on the \chf\ of the \VGpp\ process given by Proposition \ref{prop:chfVG++}.
\end{itemize}
In this first analysis we select the set of parameters reported in Table \ref{tbl:parameters}. Nevertheless, we carried out tests with different parameter sets getting similar results which we do not report here for the sake of brevity. We use the MC technique with $10^{6}$ simulations and we impose $\beta = \left(1-a\right)\alpha$ in order to have $\E\left[Z_{a}(t)\right]=t$. As far as the computation with the closed formula \eqref{eqn:optionpricingFormula} is concerned, we fix a cut-off rule for the computation of the infinite sum, namely we truncate the sum as soon as its $(n+1)$-th term contributes less than  $0.01\%$ to the sum up $n$. Finally, we model the risky asset process $F=\left\{F(t);t \ge 0\right\}$ as in Equation \eqref{eq:market:ct}.   

\begin{table}[ht]
	\scriptsize
	\centering 
	\begin{tabular}{cccccc}
		\toprule
 		$F_{0}$ & $r$ & $\sigma$ & $\theta$ & $a$ & $\alpha$ \\ [0.5ex]
		\midrule 
		100 & 0.01 & 0.2 & -0.1436 & 0.5 & 10  \\[1ex] 
		\bottomrule
	\end{tabular}
	\caption{Set of parameters we used for the numerical experiment.} 
	\label{tbl:parameters} 
\end{table}

\begin{figure*}
	\centering
	\includegraphics[scale=0.3]{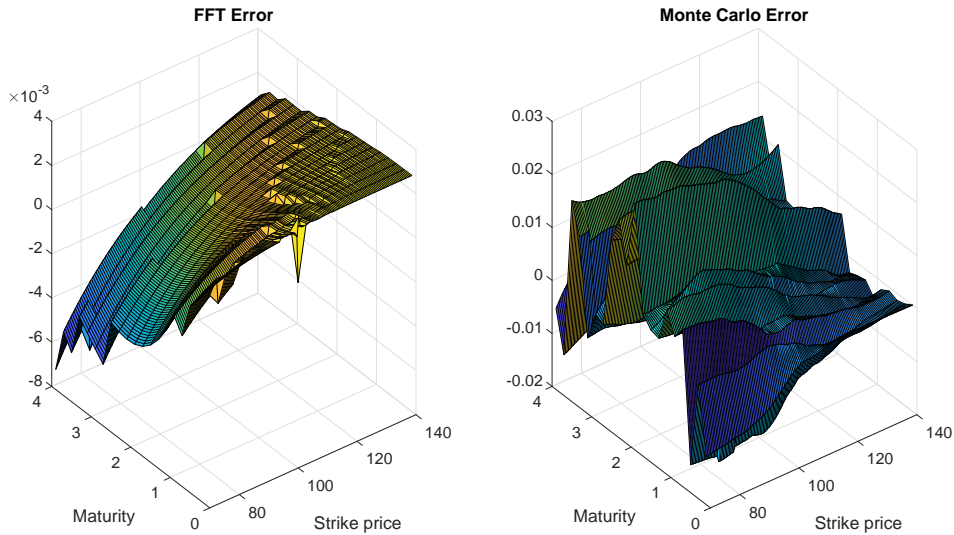}
	\caption{Fourier and MC methods error for different values of the maturity $T$ and of the strike price $K$.}
	\label{fig:MCFFTerrors}
\end{figure*}

In Figure \ref{fig:MCFFTerrors} we graphically compare the difference (error in the figures) of the FFT and MC methods with respect to the closed formula of the European call option varying the strike price $K$ and the maturity $T$. 
The size of the error of the FFT algorithm is approximately $10^{-3}$ and is smaller than that of the MC scheme  which is around $10^{-2}$. Indeed, due to its accuracy and efficiency, the FFT method is preferable for  standard contracts, whereas the MC approach is more appropriate for the pricing of more exotic derivatives.

\subsection{Calibration}
\label{subsec:Calibration}
In this subsection we show how to calibrate the \VGpp\ model on real market observations and find the set of unknown parameters $\Theta  = \left(\theta,\sigma,\alpha,a \right)$\footnote{Note that parameter $\beta$ does not appear because we imposed $b=\left(1-a\right)\alpha$ such that $\E\left[Z(t)\right]=t$.}. The data-set we rely upon is the following:

\begin{itemize}
	\item Market quotations from 23 August 2017 to 12 November 2019 of the German, Italian and Spanish power future Calendar 2020.
	\item Call options written on the German, Italian and Spanish power future Calendar 2020 with settlement date 19 November 2019 and expiration date on 13 December 2019.
	\item The risk-free rate is assumed to be $r=0.015$.
\end{itemize}

\par We perform the historical calibration with a \MLE\ relying on the closed form of the transition density of the \VGpp\ process given by Proposition \ref{prop:pdfofVgppProcess} and then numerically maximize the log-likelihood $\log \mathcal{L}\left(\Theta\right)$ with respect to $\Theta$.

\par On the other hand, one could also adopt the \emph{Generalized Method of Moments} (GMM) and minimize \enquote{a distance} between theoretical moments and their empirical analog, with respect to $\Theta$. 
Therefore, the \GMM\ method can be easily applied, by using Proposition \ref{prop:cumulants} recalling that the first cumulant is the mean, the second one is the variance and that skewness $s\left(X\right)$ and kurtosis $k\left(X\right)$ can be derived from higher order cumulants as follows:
\begin{equation*}
	s\left(X\right)=\frac{c_{3}\left(X\right)}{c_{2}\left(X\right)^{3/2}}, \qquad k\left(X\right)=\frac{c_{4}\left(X\right)}{c_{2}\left(X\right)^{2}}.
\end{equation*}     

\par
\begin{figure*}
	\centering
	\includegraphics[scale=0.3]{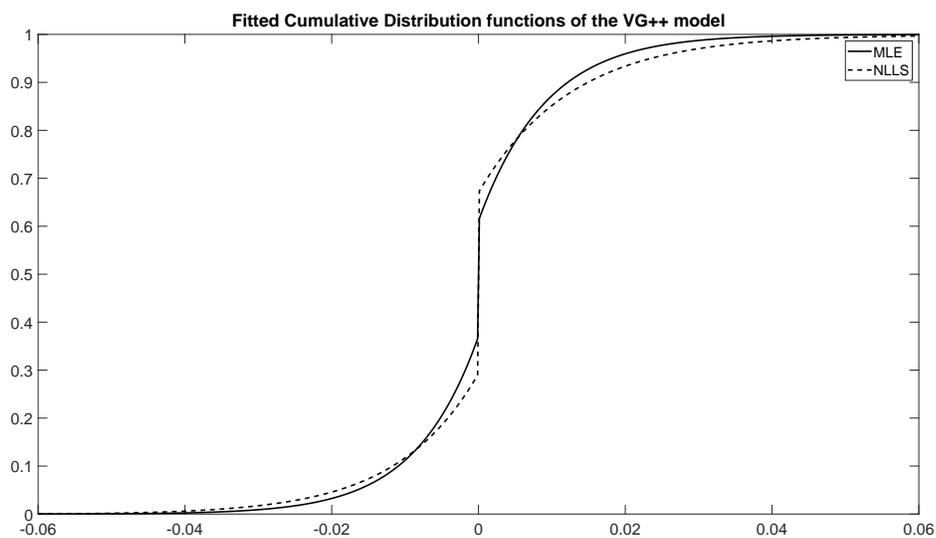}
	\caption{Fitted cumulative distribution functions of the \VGpp\ process obtained at maturity $T$ using the \MLE\ and \NLLS\ methods on Italian power forward quotations.}
	\label{fig:fittedCDF}
\end{figure*}

 The historical calibration is generally suitable for risk-management purposes, while instead the  calibration on option quotes must be considered in order to properly price derivative contracts (see \citet{CT2003}). If the market quotes $n$ products \footnote{Usually, European Call or Put options are quoted and liquid for many markets whereas more complex derivatives are traded over the counter (OTC).} 
$\left\{C_{i}\right\}_{i=1}^{n}$, the goal is then to find the set of parameters $\Theta^{*}$ which minimizes the following quantity:
\begin{equation*}
	\Theta^{*}=\argmin_{\Theta} \sum_{i=1}^{n}\left(C_{i} - C_{i}\left(\Theta\right)\right)^{2},
\end{equation*}    
where $C_{i}\left(\Theta\right)$ is the price obtained by using the \VGpp\ model. 
The optimization problem consists in a numerical Non-Linear-Least-Squared (\NLLS) problem. 
In Table \ref{tbl:VGppfittingITA}, Table \ref{tbl:VGppfittingDE} and Table \ref{tbl:VGppfittingES} we report the parameters obtained per each country with the historical calibration (\MLE) and with the calibration of option quotes (\NLLS)\footnote{For brevity we focus on the \MLE\ method and do not use the \GMM.}, whereas in Figure \ref{fig:fittedCDF} we draw the cumulative distribution functions of the \VGpp\ process at maturity $T$\footnote{Note that the density has a non-zero mass at point $x=0$.}.  \\
\par
\begin{figure*}
	\centering
	\includegraphics[scale=0.3]{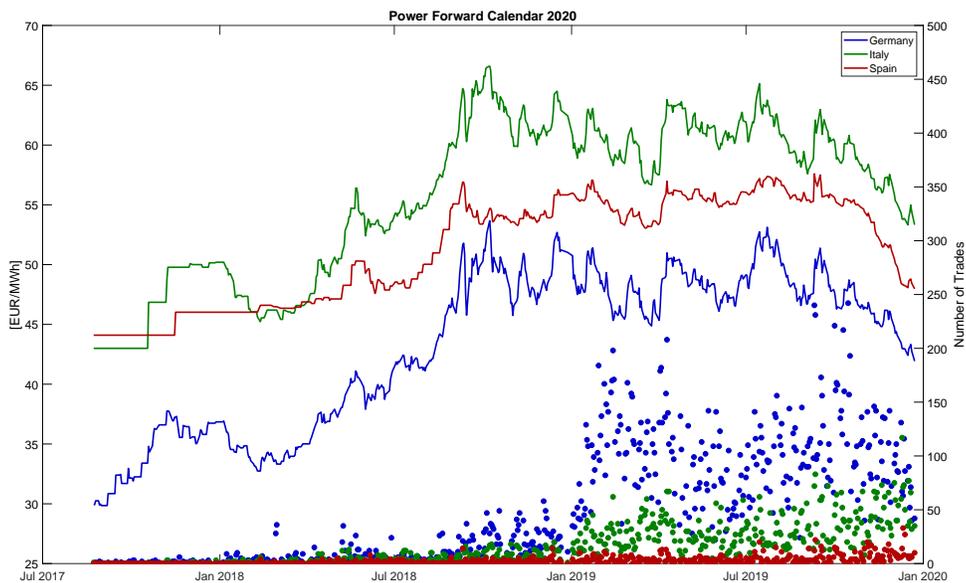}
	\caption{Prices of the German, Italian and Spanish power forward Calendar 2020 and their respective number of trades.}
	\label{fig:mktliquidity}
\end{figure*}

 European power future markets are not always liquid and, in some cases, prices tend to remain constant over time. As is shown in Figure \ref{fig:mktliquidity} the power future calendar 2020 is not very liquid, especially when the delivery is far out but its liquidity increases as the delivery approaches. For these reasons, power future markets offers a natural setting to test our model.
Indeed, the value of the parameters $a$ and $\alpha$ can be interpreted as the \emph{liquidity activity} of the market. Taking the change $\Delta X = X(t) - X(t-1)$ of the log-price over the time interval $\Delta t$, from Equation \eqref{eqn:VG++Density} we observe that the probability that the increment equals zero over the time interval $\Delta t$ is strictly larger than zero and, more precisely, it is given by
\begin{equation*}
	\mathbb{P}\left(\Delta X=0\right) = a^{\alpha \Delta t},
\end{equation*} 
since the density of the \VGpp\ process has an atom in zero. This is the main financial difference from the standard VG process which does imply that non-zero trading activity takes place in every time interval. Nevertheless, our model inherits the mathematical tractability of the standard VG process which is in any case recovered when $a$ tends to zero. 

In financial markets the liquidity is strictly related to the amount of registered transactions: if the number of trades is high, the prices fluctuate faster than when a small number of contracts is exchanged. In the extreme case where no products are traded the price remains constant over time, once again this feature cannot be captured by Brownian subordination where the subordinator has infinite activity. Therefore, illiquid markets are characterized by high values of the probability $\mathbb{P}\left(\Delta X=0\right)$. We remark once again that since the transition density of the Variance Gamma process is atom-less, such a process always presents a non zero increment over the time period $\Delta t$ and hence their paths cannot be constant over time. 

The results reported in Table \ref{tbl:VGppfittingITA}, Table \ref{tbl:VGppfittingDE} and Table \ref{tbl:VGppfittingES} are coherent with some empirical facts observed in power markets: first of all, future products are more liquid than the corresponding options: this is clear if we compare the values of $\mathbb{P}\left(\Delta X=0\right)$ obtained calibrating the model on historical forward quotations (\MLE) with the ones we get when we calibrate it on European option prices (\NLLS). Moreover, as a matter of fact, the German power future market is more liquid than the Italian and Spanish ones, as it can be observed in Figure \ref{fig:mktliquidity}: the number of trades in German future power markets is significantly higher than the one we observe in the other markets. This empirical evidence is coherent with the value of $\mathbb{P}\left(\Delta X=0\right)$ we estimate for the three markets: such a probability is smaller in the German power market than in the other ones. Finally, the Spanish market is the most illiquid one, as it can be deduced observing the number of trades in Figure \ref{fig:mktliquidity}: consequently, the values of $\mathbb{P}\left(\Delta X=0\right)$ in Table \ref{tbl:VGppfittingES} are significantly higher than the ones reported in Table \ref{tbl:VGppfittingITA} and Table \ref{tbl:VGppfittingDE}.

\begin{table}[!htb]
	\scriptsize
	
	\begin{minipage}{1\linewidth}
		\centering
	\begin{tabular}{cccccc}
	\toprule
	Method & $\sigma$ & $\theta$ & $a$ & $\alpha$ & $\mathbb{P}\left(\Delta X=0\right)$ \\ [0.5ex]
	\midrule 
	\MLE\ & 0.16 & 0.18 & 0.46 & 1255.7 & 0.02  \\
	\NLLS\ & 0.20 & 0.0.39 & 0.54 & 650.71 & 0.21  \\[1ex] 
	\bottomrule
\end{tabular}
\caption{Set of parameters $\Theta$ for the power Italian market.} 
\label{tbl:VGppfittingITA}
\vspace{0.5cm}

\begin{tabular}{cccccc}
	\toprule
	Method & $\sigma$ & $\theta$ & $a$ & $\alpha$ & $\mathbb{P}\left(\Delta X=0\right)$ \\ [0.5ex]
	\midrule 
	\MLE\ & 0.24 & 0.02 & 0.27 & 872.83 & 0.01 \\
	\NLLS\ & 0.28 & 0.91 & 0.52 & 1044.43 & 0.06 \\[1ex] 
	\bottomrule
\end{tabular}
\caption{Set of parameters $\Theta$ for the German power future market.} 
\label{tbl:VGppfittingDE}

\vspace{0.5cm}
\begin{tabular}{cccccc}
	\toprule
	Method & $\sigma$ & $\theta$ & $a$ & $\alpha$ & $\mathbb{P}\left(\Delta X=0\right)$ \\ [0.5ex]
	\midrule 
	\MLE\ & 0.09 & 0.05 & 0.38 & 6430.06 & 0.08 \\
	\NLLS\ & 0.13 & 0.83 & 0.49 & 616.35 & 0.18 \\[1ex] 
	\bottomrule
\end{tabular}
\caption{Set of parameters $\Theta$ for the power Spanish market.} 
\label{tbl:VGppfittingES}
\end{minipage}
\end{table}

\subsection{Pricing of exotic derivatives}
\label{subsec:pricing}
\par
\begin{figure*}
	\centering
	\includegraphics[scale=0.3]{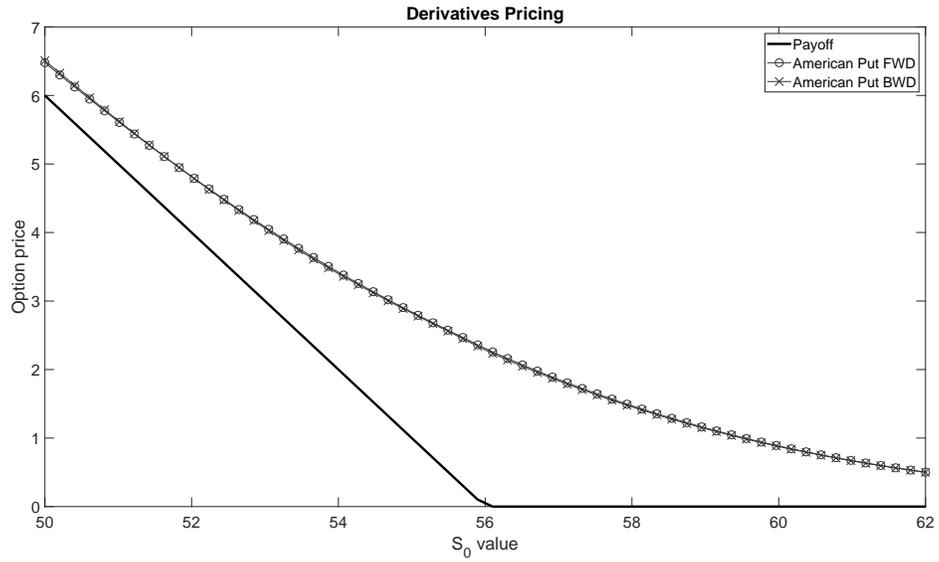}
	\caption{Price of the American Put option with different values of starting point $F(0)$ using Least-Square Monte Carlo with forward and backward simulations.}
	\label{fig:AmericanPutVGpp}
\end{figure*}

 Once that the \VGpp\ model is calibrated on quoted derivatives, it is possible to price illiquid contingent claims in a consistent way. For illustrative purposes we price American put options written on the Italian power future calendar with the Least-Square Monte Carlo introduced by \citet{LSW01} combined with the backward simulations described in Section \ref{sec:backwardsim}
and for completeness, with the sequential (forward) simulation approach. The results are reported in Figure \ref{fig:AmericanPutVGpp}, where we fix the strike price $K=56$ and the maturity $T=0.26$ years and we set different values of the process $F$ at time $t=0$. As observed, for example, in \citet{seydel2004}, the value of the American put options is never lower than the payoff and, as expected, the sequential simulation and the backward simulation return indistinguishable results. This result is not surprising, since the interpretation of the index set $I=\left\{t \ge 0\right\}$ of the stochastic process $X$ as time is just a convention: the mathematical object $X=\left\{X(t);t \in I\right\}$ is well defined even if the index set $I$ has not an order relation.
A simple question then arises: is there any advantage in using backward simulations instead of the standard forward approach? Backward simulations are not necessarily faster than forward simulations as observed in \citet{Sabino20}: nevertheless, the backward recursion of the stochastic optimization at each time step $t_{j}$ requires  the path simulations at time $t_{j}$ and $t_{j+1}$ only, which is perfectly consistent with backward approach in contrast, with the forward strategy one has to store the entire set of paths. For example, using the standard forward simulation approach to price an American contract with maturity one year, daily early exercise and $10^6$ simulations, $2.52 \cdot 10^{8}$ values need to be  stored instead of $2 \cdot 10^{6}$ values which are necessary with the backward simulations strategy. This gives a remarkable computational advantage especially if the contract has a large maturity or if one deals with the pricing of more complex derivatives such as gas storages (\citet{BDJ08}) or virtual power plants (\citet{TB2000}), for which additional discretization grids are needed.\\
\par 
In order to point out differences between the Variance Gamma and the \VGpp\ processes we apply them to the same market framework: to this aim, we consider the pricing of Lookback call options with MC simulations. We stress out once again that the transition density of the \VGpp\ process has an atom at zero and then the interval $\Delta X$ in the log-price over the time interval $\Delta t$ can be zero with strictly positive probability: this is equivalent to say that no trades have been exchanged over that time interval. On the other hand, in the Variance Gamma model a zero trading activity is not possible over any finite time interval $\Delta t$. This difference between the two models has an impact on derivative valuation. Indeed, from a financial perspective, whenever an agent sells derivatives, a hedging strategy has to be implemented. If the underlying asset is not liquid, such a hedging strategy, a delta-hedging for example, might be expensive and hard to implement. 

Indeed, if an option seller decides to adopt the delta-hedging strategy it may happen that the underlying asset is not available  therefore, the strategy can not be implemented at all. On the other hand, if the underlying asset is exchanged but the bid-ask spread is extremely wide, the hedging strategy will be highly expensive. For these reasons,  the price of options in illiquid markets should be higher than that of the same contingent claim traded in a liquid market: the price of the contingent claim must take into account the cost of the \enquote{impracticable} hedging strategy.

\begin{figure*}
	\centering
	\includegraphics[scale=0.3]{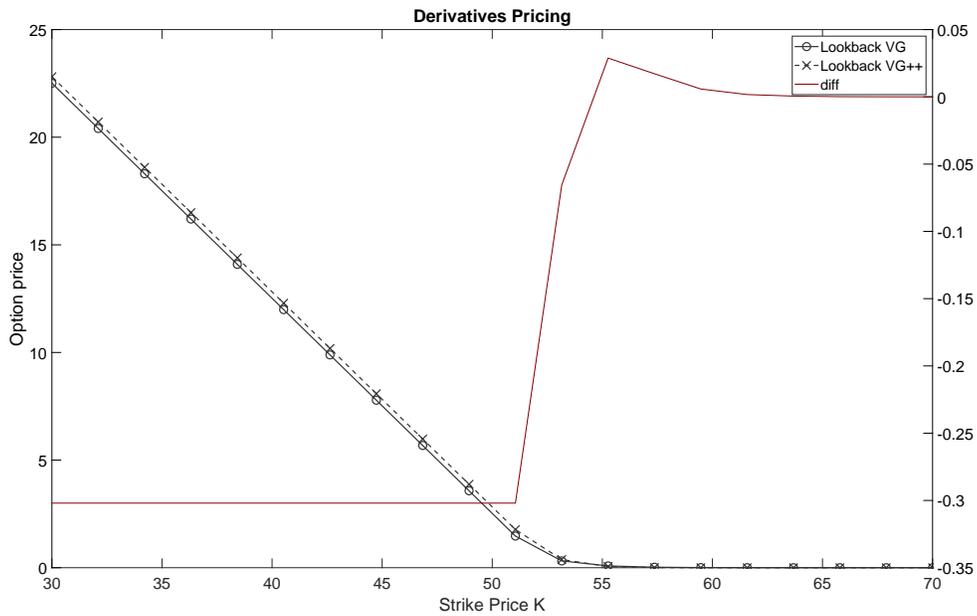}
	\caption{Price of Lookback Call option over the maximum in the Spanish market. The prices are computed using the Variance Gamma model and the \VGpp\ model, calibrated on the same data-set of vanilla options.}
	\label{fig:mlookback}
\end{figure*}

In Figure \ref{fig:mlookback} we show the price of Lookback call options on the maximum in the Spanish future market, which is the most illiquid one of the markets we analyzed. It is worth noting that the value of the option computed with the Variance Gamma model is lower than the one we obtain using the \VGpp\ model. 
As stated before, unlike the Variance Gamma model, the \VGpp\ considers the possibility that the market becomes illiquid leading to possible difficulties in the implementation of an adequate hedging strategy. Accordingly, when the market is illiquid, in order to mitigate his risk exposure, the only thing that the option seller can do is to increase the option value. We finally observe that the price differences in Figure \ref{fig:mlookback} might not seem remarkable: indeed, even if the Spanish future market has  $8\%$ of probability of not being liquid on a given day, such a level of liquidity guarantees to the option seller  to secure himself against derivative price fluctuations.
\par We conclude that, when we consider illiquid markets, the \VGpp\ model is a better choice because it allows the option seller to include in the option price a sort of \enquote{cost of market illiquidity}, which somehow mitigates the risk of not having a proper hedging strategy. 

\section{The Multivariate framework}
\label{sec:multivariateframework}
One of the most challenging tasks in financial modeling is the extension of continuous time \Levy\ models from a univariate to a multivariate framework. In the Gaussian settings, as the one proposed by \citet{BLS1973} or \citet{HJM1992}, the extension is easy since the whole dependence structure is caught by the covariance matrix. Multi-asset versions of commonly used \Levy\ models have been proposed by \citet{Buchman2017, Buchman2020Calib, Buchman2019}, \citet{Michaelsen2018} and \citet{Micha2020} among the others. Moreover, in a series of paper, \citet{Semeraro2008}, \citet{SL2010} and \citet{BB2013} presented multivariate versions of Variance Gamma and Normal Inverse Gaussian models: their results are based on the fact that the sum of random variables with a gamma (inverse Gaussian) law still has a gamma (inverse Gaussian) law if the parameters are properly chosen. Those models have been recently extended in \citet{gardini2020, gardini2021} adding a particular market feature called \textit{stochastic delay}.
As observed by \citet{SabinoPetroni2020GammaRelated}, it is worth noting that the scaling and summation properties of the Gamma laws also hold for their \arem's, namely:
\begin{itemize}
	\item If $Z_{a} \sim \Gamma^{++}\left(a, \alpha,\beta\right)$ for every $c>0$ it results:
	\begin{equation}
		cZ_{a} \sim \Gamma^{++}\left(a, \alpha,\frac{\beta}{c}\right).
		\label{eqn:gammppScaling}
	\end{equation}
\item If $Z_{a,i}\sim \Gamma^{++}\left(a, \alpha_{i},\beta\right)$, $i=1,\dots,n$ and are independent then:
\begin{equation}
	\sum_{i=1}^{n}Z_{a,i} \sim \Gamma^{++}\left(a, \sum_{i=1}^{n}\alpha_{i},\beta\right).
	\label{eqn:gammppSumming}
\end{equation}
\end{itemize}
For this reason, the same construction proposed by \citet{Semeraro2008}, \citet{SL2010} and \citet{BB2013}
can be used to build a multivariate subordinator $\boldsymbol{H} = \left\{\left(H_{1}\left(t\right),\dots,H_{n}\left(t\right)\right);t\ge 0\right\}$ whose marginal distributions have a $\Gamma^{++}$ law with suitable parameters. The construction is the following: consider independent $X_{i} = \left\{X_{i}\left(t\right); t \ge 0\right\}$ for $i=1,\dots,n$ with $X_{i}(t)\sim \Gamma^{++}\left(a,\alpha_{i}t,\frac{\beta}{c_{i}}\right)$ and consider $\zapp$ defined in Section~\ref{sec:preliminaries}. We define the process $\boldsymbol{H}$ as:
\begin{equation*}
	H_{i}(t) = X_{i}(t) + c_{i}\zapp(t), \qquad i=1,\dots,n.
\end{equation*}
where $c_{i}>0$ for all $i=1,\dots,n$. Using properties \eqref{eqn:gammppScaling} and \eqref{eqn:gammppSumming} it is easy to check that $H_{i}\left(t\right) \sim \Gamma^{++}\left(a, \left(\alpha_{i}+\alpha\right)t,\frac{\beta}{c_{i}}\right)$. All the components of the process $\boldsymbol{H}$ are dependent, because of the presence of the common process $\zapp$. $\boldsymbol{H}$ is a multivariate subordinator and it can be used to derive multidimensional versions of the \VGpp\ process: this topic will be the subject of future investigations.

\section{Conclusions and future inquires}
\label{sec:conclusions}
In this paper we have introduced a new \Levy\ process, named  Variance Gamma++, which inherits both the mathematical tractability and the financial interpretation of Variance Gamma process. Such a new process,  has an additional parameter which can be interpreted as a measure of the market liquidity. 
\par The construction is based on a time-changed Brownian motion, where the time-change is given by a subordinator which is derived from the self-decomposability of the gamma law. Using the results in \citet{petronisabino2020} we have given the full characterization of this subordinator in terms of its \Levy\ triplet, accordingly have found the one of the Variance Gamma++ and finally have proven that the latter process is of finite activity and of finite variation.
\par Unlike the Variance Gamma process, whose transition density does not present an atom at the origin, it turns out that the Variance Gamma++ process allows null increments in any finite time interval. For this reason, the Variance Gamma++ is a good candidate to model illiquid markets, in which prices tends to be constant over time, and preserves at the same time, all the strengths of the Variance Gamma, namely a closed form pricing formula for vanilla options and an explicit expression both for characteristic function and transition probability density. In particular, the evaluation of the closed formula for European options  does not require the numerical computation of any integral and hence turns out to be extremely efficient from the computational point of view.
\par Moreover, we have derived algorithms for the forward and the backward simulation of the skeleton of subordinator and of the Variance Gamma++ process. The backward simulation approach is instrumental to price American derivative contracts and has the advantage of avoiding to store  the whole set of trajectories, leading to a remarkable saving of the RAM memory space.
\par We have shown that the Variance Gamma++ is particularly appropriate to model illiquid markets and have applied it to future power markets, which usually presents periods of low liquidity.  To this end, we have calibrated the new Variance Gamma++ process on real data using both the \MLE\ and the \NLLS\ techniques. Consequently, we have priced exotic derivatives and we have highlighted the differences with the original Variance Gamma process. In particular,  our model tends to return higher prices for derivatives in illiquid markets than the Variance Gamma model. This is  expected from a financial point of view, since in illiquid markets the hedging strategies are difficult to implement and therefore, option sellers tend to increase the option premia.
\par In addition, we have illustrated how to extend the Variance Gamma++ process to a multidimensional framework, following the approach proposed by \citet{Semeraro2008}, \citet{SL2010} and \citet{BB2013} whereas, concrete applications will be the subject of future inquires. 
\par Finally, a topic deserving further investigation is the possibility to use the procedure adopted to construct the Variance Gamma++ process to the inverse Gaussian law, which is a self-decomposable distribution as well, and accordingly study its mathematical properties and potential financial applications.

\clearpage
\begin{appendices}

\section{Variance Erlang distribution: derivation and option pricing}
In this Appendix we report some results about Exponential Polynomial Trigonometric (EPT) distributions we used in the article. For a complete discussion about this topic refer to \citet{SextonHanzon12}.
\subsection{2-EPT distributions}
The class of EPT functions $f:\left[0,\infty\right) \to \R$ is given by:
\begin{equation*}
	f\left(x\right) = \Re\left(\sum_{k=1}^{K}p_{k}\left(x\right)e^{\mu_{k}x}\right)
\end{equation*}
where $\Re\left(z\right)$ denotes the real part of a complex number $z\in \mathbb{C}$, $p_{k}\left(x\right)$ is polynomial with complex coefficients for each $k=1,\dots,K$ and $\mu_{k}\in \mathbb{C}$ for $k=1,2,\dots,K$. And EPT function defined on the positive real line can be represented in the following form:
\begin{equation*}
	f\left(x\right) = \boldsymbol{c}e^{\boldsymbol{A} x} \boldsymbol{b}, \qquad x\ge 0,
\end{equation*}
where $\boldsymbol{A}$ is a $n\times n$ matrix, $\boldsymbol{c}$ is $1\times n$ vector and $\boldsymbol{b}$ is a $n\times 1$ vector. We consider probability density functions which can be written as two separate EPT functions:

\begin{equation*}
	f\left(x\right) =
	\begin{cases} 
		\boldsymbol{c}_{N}e^{\boldsymbol{A}_{N} x} \boldsymbol{b}_{N}, \qquad x\ge 0,\\
		\boldsymbol{c}_{P}e^{\boldsymbol{A}_{P} x} \boldsymbol{b}_{P}, \qquad x>0.
	\end{cases}
\end{equation*}

\subsection{Variance Gamma as an 2-EPT distribution}
The Variance Gamma law can be viewed as an 2-EPT distribution under some parameter constrains. Its \pdf\ and \chf\ are given by:
\begin{equation*}
f_{X}\left(x;C,G,M\right) = \frac{\left(GM\right)^{C}}{\sqrt{\pi}\Gamma\left(C\right)}\exp\left(\frac{\left(G-M\right)x}{2}\right)\left(\frac{|x|}{G+M}\right)^{C-\frac{1}{2}}K_{C-\frac{1}{2}}\left(\frac{\left(G+M\right)|x|}{2}\right)
\end{equation*}
\begin{equation*}
	\phi_{X}\left(u\right)= \left(\frac{GM}{GM + \left(M-G\right) iu + u^2}\right)^{C}.
\end{equation*}
where $K_\nu(z)$ denotes the modified Bessel function of the second kind and $C,G,M \in \R^{+}$.
Following \citet{SextonHanzon12} we show that the Variance Gamma law is an 2-EPT distribution if $C\in \N$. According to \citet[pag. 443]{abramowitzstegun1964} we have:
\begin{equation*}
	\sqrt{\frac{\pi}{2x}}K_{n+\frac{1}{2}}\left(x\right)= \left(\frac{\pi}{2x}\right)e^{-x}\sum_{k=0}^{n}\left(n+\frac{1}{2},k\right)\left(2z\right)^{-k},
\end{equation*}
where 
\begin{equation*}
	\left(n+\frac{1}{2},k\right) = \frac{\left(n+k\right)!}{k!\Gamma\left(n-k+1\right)},
\end{equation*}
therefore after some algebra, $f_X(x)$ can be rewritten as
\begin{equation*}
	f_{X}(x)  = \exp\left(\frac{\left(G-M\right)x}{2} - \frac{\left(G+M\right)|x|}{2}\right)           \frac{\left(GM\right)^{C}}{\left(C-1\right)!}\sum_{k=0}^{C-1} \frac{\left(C-1+k\right)!\left(G+M\right)^{-C-k}|x|^{C-1-k}}{\left(C-1-k\right)!k!}.
\end{equation*}
We can split the density around the origin, obtaining:
\begin{equation}
	f_{X}\left(x\right) =
	\begin{cases} 
		\exp\left(Gx\right) \frac{\left(MG\right)^{C}}{\left(C-1\right)!}\sum_{s=0}^{C-1}\frac{\left(2\left(C-1\right)-s\right)!\left(G+M\right)^{-2C+1+s}|x|^{s}}{s!\left(C-1-s\right)!},
		 &\qquad x\le 0,\\
	\exp\left(-Mx\right) \frac{\left(MG\right)^{C}}{\left(C-1\right)!}\sum_{s=0}^{C-1}\frac{\left(2\left(C-1\right)-s\right)!\left(G+M\right)^{-2C+1+s}|x|^{s}}{s!\left(C-1-s\right)!},
	&\qquad x> 0.
	\end{cases}
\label{eqn:vgasEPT}
\end{equation}
Observe that the polynomial parts of \eqref{eqn:vgasEPT} are identical for all $x$ and this implies that $\boldsymbol{c}_{N} = \boldsymbol{c}_P$ and $\boldsymbol{b}_{N} = \boldsymbol{b}_{P}$. We set:
\begin{align*}
\boldsymbol{c} &= \left(c_{0},\dots,c_{S-1}\right), & \boldsymbol{c} \in \R^{1 \times C}\\
c_{s} &= \frac{\left(MG\right)^{C}}{\left(C-1\right)!}\frac{\left(2\left(C-1\right)-s\right)!\left(G+M\right)^{-2C+1+s}}{\left(C-1-s\right)!}, & s \in \left(0,\dots,C-1\right).
\end{align*}
Similarly $\boldsymbol{b} = \left(1,0,\dots,0\right)^{T}$ is a $C\times 1$ column vector whereas $\boldsymbol{a}$ is given by:

\begin{equation*}
	\boldsymbol{a}= 
	\begin{pmatrix}
		0& 0& \cdots & 0&0 \\
		1& 0& \cdots &0& 0\\
		0 & 1 & 0 \cdots & 0 & 0\\
		\vdots  & \vdots  & \ddots & \vdots & \vdots  \\
		0& 0& \cdots & 1 &0
	\end{pmatrix},
\end{equation*}
and finally we get that $p\left(x\right) = \boldsymbol{c}e^{-\boldsymbol{a}x}\boldsymbol{b}$. Summarizing, we have:
\begin{equation*}
	f_{X}\left(x;C,G,M\right) =
\begin{cases} 
\boldsymbol{c} e^{Gx} e^{-\boldsymbol{a}x} \boldsymbol{b}  & x\le 0,\\
	\boldsymbol{c} e^{-Mx} e^{\boldsymbol{a}x} \boldsymbol{b} & x> 0.
\end{cases}
\end{equation*}

\noindent Finally, defining $\boldsymbol{A}_{N} = G \boldsymbol{I}-\boldsymbol{a}$ and $\boldsymbol{A}_{P} = -M\boldsymbol{I} + \boldsymbol{a}$, the \pdf\ of a Variance Gamma law with $C\in\N$ results:
\begin{equation*}
	f_{X}\left(x;C,G,M\right) =
	\begin{cases} 
		\boldsymbol{c} e^{\boldsymbol{A}_{N}x} \boldsymbol{b}  & x\le 0,\\
		\boldsymbol{c} e^{\boldsymbol{A}_{P}x}  \boldsymbol{b} & x> 0.
	\end{cases}
\end{equation*}

\subsection{The price process}
We model the risky underlying asset $F$ as:

\begin{equation*}
	F(t) = F(0) e^{rT + \omega T + X(T)},\quad F(0) = F_{0}
\end{equation*}
where $T\ge 0$, $r$ is the risk-free rate and $\omega$ is such that the discounted price process is a martingale. In order to work under the risk-neutral measure $\mathbb{Q}$ we must require that:

\begin{equation*}
	\E^{\mathbb{Q}}\left[e^{\omega T + X(T)}\right]= 1
\end{equation*}
and this leads to:
\begin{equation*}
	\omega  = C \log\left(\left(1 - \frac{1}{M}\right)\left(1 + \frac{1}{G}\right)\right).
	\label{eqn:riskneutralityVE}
\end{equation*}
If we add the constrain $CT \in \mathbb{N}$, we observe that $\omega$ is defined only if $M >1$. Moreover, if $CT \in \mathbb{N}$ a closed formula for a Call option with maturity $T$ can be derived (In the original article you have $\tau= T-t$, which is the time to maturity, instead of $T$: here we considered $t=0$ and hence $\tau$ and $T$ coincides).

\subsection{A closed formula for Call option pricing}
\label{sec:VarianceErlangOptionFormula}
Consider a Call option with strike price $K$ and maturity $T$. The value of the underlying asset at $t=0$ is $F(0)=F_{0}$ and we consider a constant risk free rate $r\ge 0$. Define:
\begin{equation*}
	d = \log\left(\frac{F(0)}{K}\right) + \left(r+\omega\right)T.
\end{equation*} 
The price of the Call option $C(0,K)$, where $X(T)$ has a infinitely divisible distribution with 2-EPT density distribution with realizations $\left(\boldsymbol{A}_{N},\boldsymbol{b}_{N},\boldsymbol{c}_{N},\boldsymbol{A}_{P},\boldsymbol{b}_{P},\boldsymbol{c}_{P}\right)$, is given by:
\begin{itemize}
	\item If $d >0$:
	\begin{equation*}
	\begin{split}
	C(0,K) & = F(0)e^{\omega T} \left(\boldsymbol{c}_{N}\left(\boldsymbol{A}_{N} + \boldsymbol{I}\right)^{-1}\right)\boldsymbol{b}_{P} - \boldsymbol{c}_{N}\left(\boldsymbol{A}_{N } + \boldsymbol{I}\right)^{-1} e^{-\left(\boldsymbol{A}_N + \boldsymbol{I}\right)d}\boldsymbol{b}_{N} \\
	& - \boldsymbol{c}_{P}\left(\boldsymbol{A}_{p} + \boldsymbol{I}\right)^{-1}\boldsymbol{b}_{P} - K e^{-rT} \left(1 - \boldsymbol{c}_{N}\boldsymbol{A}_{N}^{-1}e^{-\boldsymbol{A}_{N}d}\boldsymbol{b}_{N}\right).
		\end{split}
	\end{equation*}
\item If $d\le 0$:
\begin{equation*}
	C(0,K) = -F(0)e^{\omega T} \boldsymbol{c}_{P} \left(\boldsymbol{A}_{P} + \boldsymbol{I}\right)^{-1}e^{-\left(\boldsymbol{A}_{p} + \boldsymbol{I}\right)d}\boldsymbol{b}_{p} + K e^{-rT} \boldsymbol{c}_{P} \boldsymbol{A}_{P}^{-1} e^{-\boldsymbol{A}_{p}d}\boldsymbol{b}_{P}.
\end{equation*}
\end{itemize}
In contrast to many option pricing formulas available in finance, observe that no integrals appear: the computation of $C\left(0,K\right)$ requires only linear algebra techniques which are usually faster than numerical integration procedures.

\subsection{From $C,G,M$ to $\alpha,\beta, \sigma,\theta$}
Usually in literature, the parametrization of the Variance Gamma process is given in term of $\alpha,\beta,\sigma$ and $\theta$, whereas in the previous section the 2-EPT version of the Variance Gamma is a function of $C,G$ and $M$. Since these equivalent parametrization may be a source of confusion, in this section we show how to easily switch from one to the other. For the sake of completeness, we recall how the Variance Gamma process is defined.
\begin{defn}
	Consider the gamma process $G = \left\{G(t);t\ge 0\right\}$ such that $G(t)\sim \Gamma\left(\alpha t, \beta\right)$ and consider a Brownian motion $W$ with drift $\theta \in \R$ and diffusion $\sigma \in \R^{+}$, independent of $G$. The process $X = \left\{X(t);t\ge 0\right\}$ defined as:
	\begin{equation}
		X(t) = \theta G(t) + \sigma W(G(t)) \quad t\ge 0,
		\label{eqn:variacegammaprocess}
	\end{equation}
	is called Variance Gamma process and its characteristic function at time $t>0$ is given by:
\begin{equation}
	\phi_{X(t)}(u) = \left(1 - \frac{i}{\beta}\left(u\theta + iu^{2}\frac{\sigma^2}{2}\right)\right)^{-\alpha t}.
	\label{eqn:chfVGprocess}
\end{equation}
\end{defn}
\vspace{0.2cm}
\noindent Observe that Equation \eqref{eqn:chfVGprocess}, can be rewritten as:
\begin{equation*}
	\phi_{X\left(t\right)} = \left(1 -\frac{1}{\beta}\left(u\theta + i\frac{\sigma^2}{2}u^{2}\right)\right)^{-\alpha T} = \left(\frac{2\frac{\beta}{\sigma^2}}{2\frac{\beta}{\sigma^2}-iu\frac{2\theta}{\sigma^2} + u^2}\right)^ {\alpha T},
\end{equation*}
that has to be compared to:
\begin{equation*}
	\phi_{X\left(t\right)}(u) = \left(\frac{GM}{GM + \left(M-G\right)iu + u^2}\right)^{C},
\end{equation*}
and hence,
\begin{align*}
GM & = 2\frac{\beta}{\sigma^2},\\
M-G & = -2\frac{\theta}{\sigma^2}.	
\end{align*}
Finally we obtain:
\begin{align*}
G & = \frac{1}{\sigma^2} \left(\theta + \sqrt{\theta^2 + \beta\sigma^2}\right),	\\
 M & = \frac{\sqrt{\theta^2 + \beta \sigma^2}}{\sigma^2} - \frac{\theta}{\sigma^2}.
\end{align*}

\end{appendices}

\clearpage


\clearpage
\bibliographystyle{plainnat}
\bibliography{biblioAll}

\end{document}